\documentclass[a4paper, 12pt]{article}
\usepackage{amsmath,amssymb,amsthm,amsfonts,epsf}
\usepackage{graphicx}% epstopdf
\usepackage{latexsym,framed,color,url}
\usepackage[margin=2cm]{geometry}
\def\b{\begin{eqnarray}}
\def\e{\end{eqnarray}}
\def\n{\noindent}
\def\bbbc{{\Bbb C}}
\def\circledS{{\,\,s\!\!\!\!\!\!\bigcirc\,\,}}
\newcommand{\bfi}{\bfseries\itshape}

\newcommand{\rem}[1]{}

\makeatletter

\@addtoreset{figure}{section}
\def\thefigure{\thesection.\@arabic\c@figure}
\def\fps@figure{h, t}
\@addtoreset{table}{bsection}
\def\thetable{\thesection.\@arabic\c@table}
\def\fps@table{h, t}
\@addtoreset{equation}{section}

\makeatother

\begin{document}

\begin{center}

{\LARGE\textbf{Two-component {CH} system:\\
Inverse Scattering, Peakons and Geometry
\\}} \vspace {10mm} \vspace{1mm} \noindent

{{\bf D. D. Holm $^{1}$} and  {\bf R. I.
Ivanov}\footnote{Department of Mathematics, Imperial College
London. London SW7 2AZ, UK.  \texttt{d.holm@imperial.ac.uk, r.ivanov@imperial.ac.uk}}
$^{,}$}\footnote{School of Mathematical Sciences, Dublin Institute of Technology, Kevin Street, Dublin 8, Ireland,  \texttt{rivanov@dit.ie}}

\vskip1cm \hskip-.3cm

%\begin{tabular}{c}
%\hskip-1cm $\phantom{R^R} ^{a}${\it School of Mathematics, Trinity
%College Dublin,}
%\\ {\it Dublin 2, Ireland} \\
%$\phantom{R^R}^{b}${\it Institute for Nuclear Research and Nuclear
%Energy,}\\ {\it Bulgarian Academy of Sciences,} \\
%{\it 72 Tzarigradsko chaussee, 1784 Sofia,
%Bulgaria} \\
%{\it Tel:  + 353 - 1 - 608 2898 }\\{\it  Fax:  + 353 - 1- 608 2282} \\
%\\{\it $^\dag$e-mail: adrian@maths.tcd.ie}
%\\{\it $^\ddag$e-mail: gerjikov@inrne.bas.bg}
%\\ {\it $^\ast$e-mail: ivanovr@tcd.ie}
%\\
%\hskip-.8cm
%\end{tabular}
%\vskip1cm
\end{center}

%\vskip1cm

\begin{abstract}
An inverse scattering transform method corresponding to a Riemann-Hilbert problem is formulated for CH2, the two-component generalization of the Camassa-Holm (CH) equation. As an illustration of the method, the multi - soliton solutions corresponding to the reflectionless potentials are constructed in terms of the scattering data for CH2.

%{\bf PACS:} 02.30.Ik, 05.45.Yv, 45.20.Jj, 02.30.Jr

%{\bf MSC:} 35P25, 35Q15, 35Q35, 35Q51, 35Q53

%{\bf Key Words:} Hamiltonian systems, Integrable Systems, Lax Pair,
%Riemann-Hilbert Problem, Solitons.

\end{abstract}

\tableofcontents

\section{Introduction}

{\bf Purpose of the paper:} \\In this paper we investigate various aspects
of the two-component CH system (CH2), including its soliton solutions in the inverse scattering framework.
The main difference from the standard inverse scattering transform method  is that the spectral problem for CH2 is a Schr\"odinger equation with an `energy dependent' potential that is quadratic in the spectral parameter. 

\subsection{CH equation}

This section introduces the CH equation and its two-component extension, CH2. Later sections will discuss the isospectral problem for the CH2 system, leading eventually to its multi - soliton solutions. 

The CH equation \cite{CH93, CHH94}
\begin{equation}\label{eq1}
 u_{t}-u_{xxt}+2\omega u_{x}+3uu_{x}-2u_{x}u_{xx}-uu_{xxx}=0,
\end{equation}
has gained popularity as an integrable model
describing the unidirectional propagation of shallow water waves
over a flat bottom \cite{CH93,CHH94, DGH03, DGH04, J02, J03a}  as well
as that of axially symmetric waves in a hyperelastic rod
\cite{Dai98}. 
In the shallow water wave interpretation of CH, the real parameter $\omega$ is the asymptotic value of the horizontal fluid velocity $u$ at spatial infinity, as $|x|\to \infty $.   
For $\omega=0$ the CH equation possesses singular solution
in the form of peaked solitons (peakons) \cite{CH93,CHH94,Ho2010}. Summaries of the developments of many results about the CH equation appear in, e.g., \cite{HoScSt2009,Ho2010,HI10} and references therein. In the present context its most important properties are its bi-Hamiltonian structure and its Lax pair.  

The CH equation in (\ref{eq1}) may be expressed in bi-Hamiltonian form as 
\begin{equation}\label{eq2}
 m_{t}
 =-(\partial-\partial^{3})\frac{\delta H_{2}[m]}{\delta m}
 =-(\partial m + m\partial + 2\omega \partial )\frac{\delta H_{1}[m]}{\delta m},
\end{equation}
where the momentum $m$ associated to the fluid velocity $u$ is given by 
\b\label{eq4a} m = u-u_{xx} \e \n 
and the two Hamiltonians are
\b \label{eq2a} H_{1}[m]&=&\frac{1}{2}\int m u dx
, \\\label{eq2b}
H_{2}[m]&=&\frac{1}{2}\int(u^{3}+uu_{x}^{2}+2\omega u^{2})dx. \e

\n The integration is taken over the real line for functions rapidly decaying as $|x|\to \infty $, and taken over one period in the periodic case. (In the periodic case, $\omega$ is related to the mean depth.) 

The CH equation admits an infinite sequence of conservation laws
(multi-Hamiltonian structure) $H_n[m]$, $n=0,\pm1, \pm2,\ldots$,
obtainable from the recursion relation 
\begin{equation}\label{eq2aa}
 -(\partial-\partial^{3})\frac{\delta H_{n}[m]}{\delta m}
 =
 -(\partial m + m\partial + 2\omega \partial )\frac{\delta H_{n-1}[m]}{\delta m}.
\end{equation}

The recursion relation for CH leads to its Lax pair. Namely, the CH equation follows as the compatibility condition for the Lax pair \cite{CH93, CHH94}
\b \label{eq3} \Psi_{xx}&=&\Big(\frac{1}{4}+\lambda
(m+\omega)\Big)\Psi
, \\\label{eq4}
\Psi_{t}&=&\Big(\frac{1}{2\lambda}-u\Big)\Psi_{x}+\frac{u_{x}}{2}\Psi+\gamma\Psi
,\e

\n in which $\gamma$ is an arbitrary constant.

\subsection{From CH to CH2}

\n An integrable two-component generalization of the CH equation can be easily obtained by extending the Lax pair for CH in (\ref{eq3}), (\ref{eq4}) to a Lax pair whose eigenvalue problem is quadratic in the spectral parameter \cite{CLZ05}

\b \label{L1} \Psi_{xx}&=&\left(-\lambda^2 \rho^2(x)+\lambda
q(x)+\frac{1}{4}\right)\Psi, \\
\Psi_{t}&=&\Big(\frac{1}{2\lambda}-u\Big)\Psi_{x}+\frac{u_{x}}{2}\Psi,
\label{L2} \e

The compatibility of the two equations (\ref{L1}), (\ref{L2}) produces a
two-component extension of the CH equation, abbreviated as CH2,

%\begin{framed}
\b q_t\!\!&+&\!\!uq_x+2q u_x + \rho \rho_x=0,\label{LL1}\\
\rho_t\!\!&+&\!\!(u\rho)_x=0,\label{LQ1}\e
%\end{framed}

where $q=u-u_{xx}+\omega$ with $\omega$ being a constant. In our further considerations, we shall assume the limit relation $\lim_{|x|\to \infty }(\rho(x)-\rho_0)=0$,
where $\rho_0>0$ is a constant, while both $u(x)$ and $\rho(x)-\rho_0$ are Schwartz class functions. Taking $\rho=\rho_0=0$ reduces the CH2 system to the CH equation.

The CH2 energy Hamiltonian is given by
\begin{equation}\label{im}
H=\frac{1}{2}\int[u^2+u_x^2+(\rho-\rho_0)^2]\text{d}x,
\end{equation} 
and is positive-definite. The CH2 system (\ref{LL1}-\ref{LQ1}) is {\bfi bi-Hamiltonian}. This means it has two compatible Poisson brackets. Its first Poisson bracket between two functionals $F$ and $G$ of the variables $m$ and $\rho$ is in {\bfi semidirect-product Lie-Poisson form} \cite{HoMaRa1998,HoGM2}
\begin{equation}\label{ipb}
\{F,G\}_1=-\int\bigg[\frac{\delta F}{\delta m}(m\partial+\partial
m)\frac{\delta G}{\delta m}
+\frac{\delta F}{\delta m}\rho\,\partial\frac{\delta G}{\delta \rho}
+\frac{\delta F}{\delta \rho}\partial \rho \frac{\delta G}{\delta m} 
\bigg]\text{d}x.
\end{equation} 
This Poisson bracket generates the CH2 system from the Hamiltonian
$H_1 =\frac{1}{2}\int (um+\rho^2)\text{d}x$ with $m=u-u_{xx}$. Its second Poisson
bracket has constant coefficients, 
\b \{F,G\}_2=-\int\Big[\frac{\delta F}{\delta
m}(\partial-\partial^{3})\frac{\delta G}{\delta m}+\frac{\delta
F}{\delta \rho}
\partial\frac{\delta G}{\delta \rho}\Big]\text{d}x,  \label{pb2} \e
and corresponds to the Hamiltonian $H_2=\frac{1}{2}\int
(u\rho^2+u^3+uu_x^2)\text{d}x$. There are two Casimirs for the second bracket: $\int \rho\, \text{d}x$ and $\int m\, \text{d}x$. Since $H$ and $H_1$ differ only by a Casimir of the second bracket, they both generate the same flow  (uniform translation: $m_t+m_x=0$ and $\rho_t+\rho_x=0$) under the Poisson bracket (\ref{pb2}).

The CH2 system represents a two-component generalization of the CH equation. It 
was initially introduced in \cite{SA} as a tri-Hamiltonian system, and was studied further by others, see, e.g., \cite{LZ05,CLZ05,F06,I06,CI08,HLT09,GL10,SMA}. The CH2 model has various applications. For example:
\begin{itemize}
\item
In the context of shallow water waves propagating over a flat bottom, $u$ can be interpreted as the horizontal
fluid velocity and $\rho$ is the water depth  in the first approximation \cite{CI08,I09}. 
\item
In Vlasov plasma models, CH2 describes the closure of the kinetic moments of the single-particle probability distribution for geodesic motion on the symplectomorphisms \cite{HT09}. 
\item
In the large-deformation diffeomorphic approach to image matching, the CH2 equation is summoned in a type of matching procedure called metamorphosis \cite{HoTrYo2009}. 
\end{itemize}

For discussions of the geometric aspects of the CH2 system we refer to \cite{HoScSt2009,K07,HoTrYo2009}.
Its analytical properties such as well-posedness and wave breaking were studied in \cite{ELY07,CI08,H09,ZY10,GL10,CL10} and elsewhere.
In general, one can show that small initial data of the CH2 system develop into global solutions, while for some initial data wave breaking occurs. Only the plus sign  ($+$) in front of the $\rho\rho_x$
term (\ref{LL1}) corresponds to a positively
defined Hamiltonian and straightforward physical applications. It would be
interesting to know whether the model with the choice of the minus sign in (\ref{LL1}) has a physical interpretation, since this case is also integrable \cite{CI08}.

\begin{figure}[t]
\begin{center}
\includegraphics*[width=0.475\textwidth]{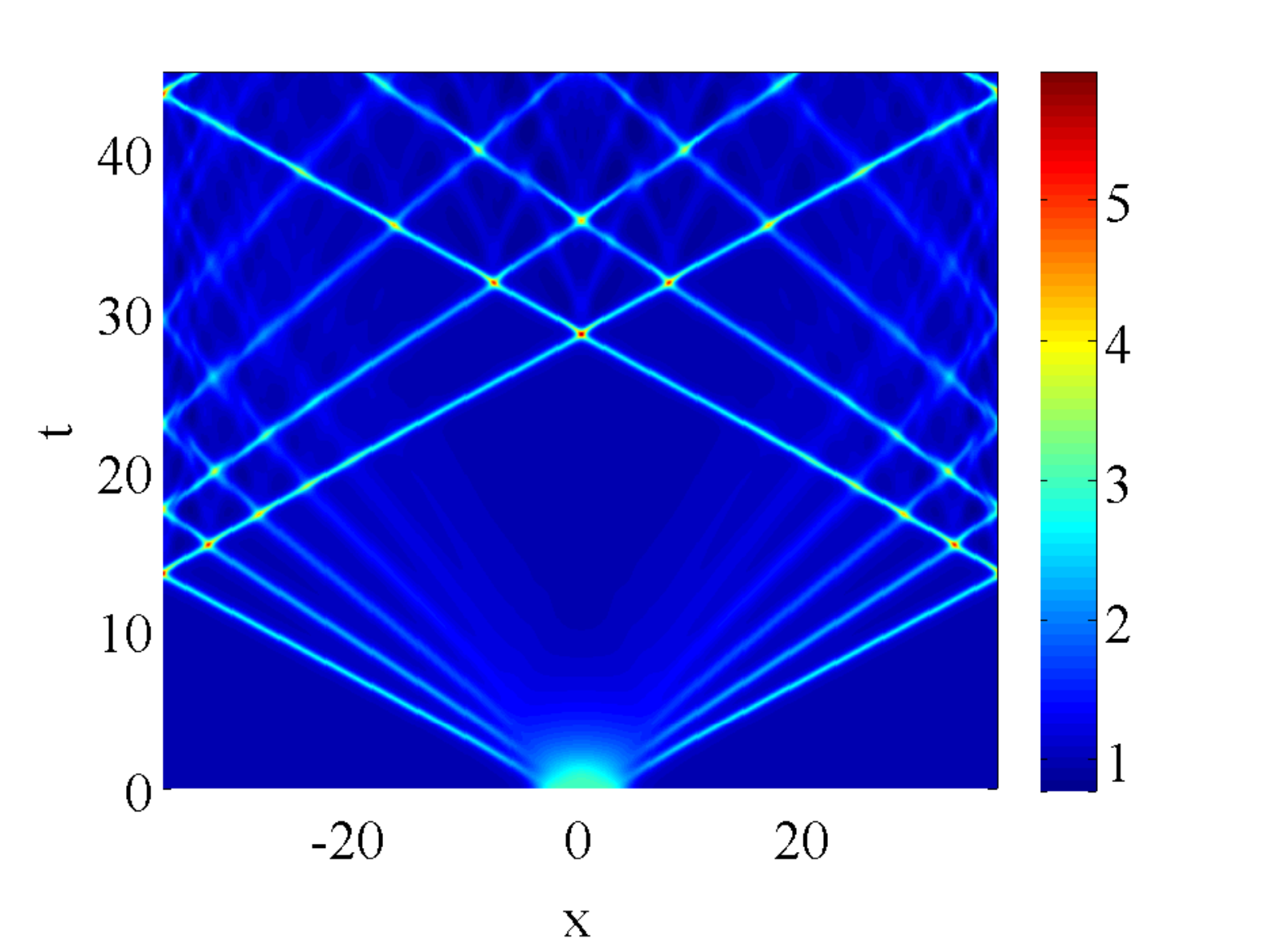}
\includegraphics*[width=0.475\textwidth]{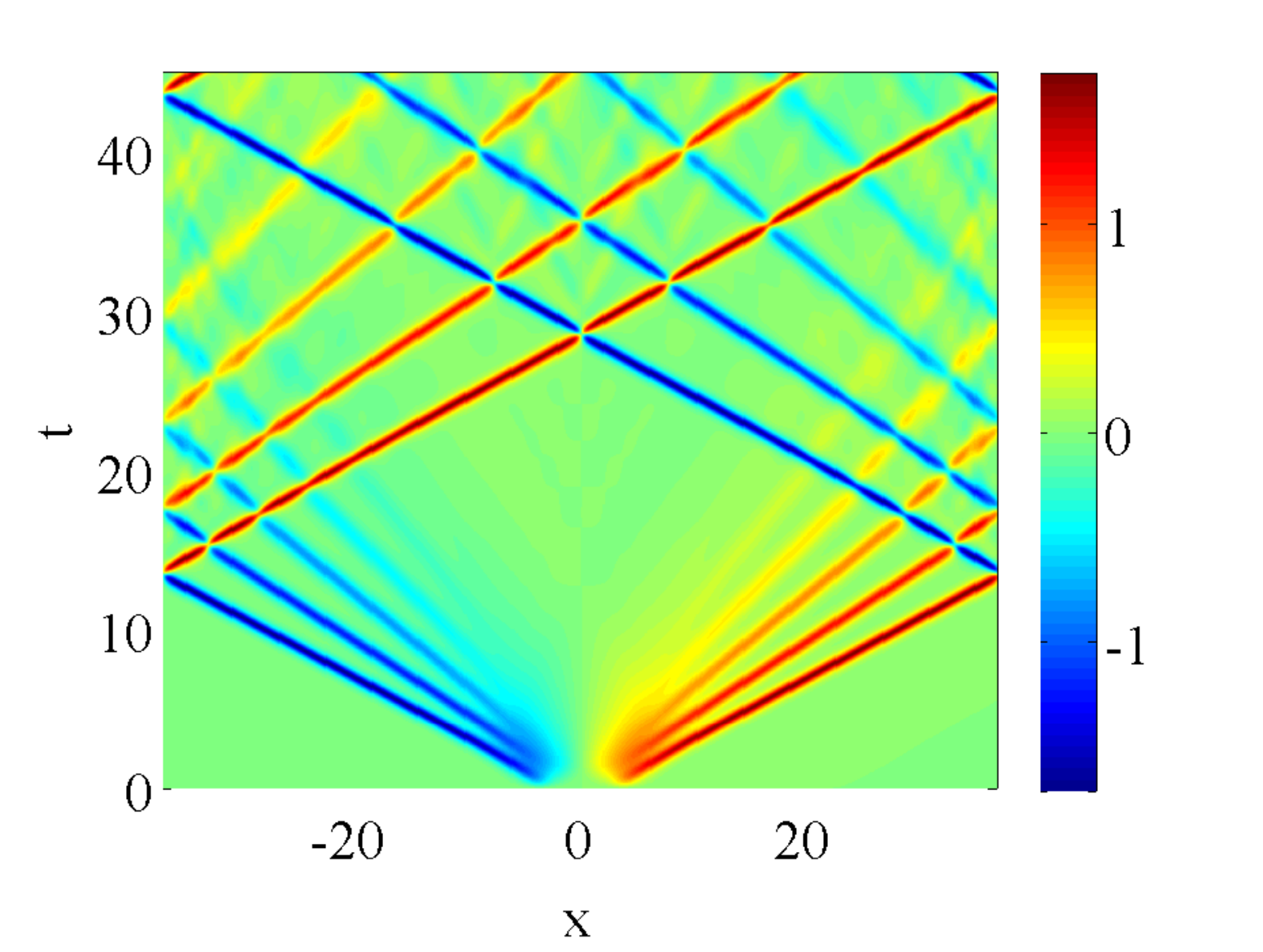}
\end{center}
\caption{\label{CH2-figs} Dam-break results for the CH2 system in equations (\ref{LL1}-\ref{LQ1}) show evolution of the density $\rho$ (left panel) and velocity $u$ (right panel), arising from initial conditions (\ref{dambreak-ic}) in a periodic domain. The color bars show positive density on the left and both positive and negative velocity on the right. The soliton solutions are seen to emerge after a finite time, and the evolution of both variables generates more and more solitons propagating in both directions as time progresses.   Figures are courtesy of L. \'O N\'araigh. } 
\end{figure}

\paragraph{Solutions of CH2 for dam-break initial conditions.}
Figure \ref{CH2-figs} plots the evolution of CH2 solutions for $(u,\rho)$ governed by equations (\ref{LL1}-\ref{LQ1}) with the $+$ sign choice in the periodic domain $\left[-L,L\right]$ with {\bfi dam-break initial conditions} given by
\begin{equation}
u\left(x,0\right)=0,\qquad{\rho}\left(x,0\right)
= 1 +  \tanh(x+a)-\tanh(x-a) 
\,,
\label{dambreak-ic}
\end{equation}
where $a\ll L$.

The dam-break problem involves a body of water of uniform depth, initially 
retained behind a barrier, in this case at $x=\pm a$.  When the barrier is
suddenly removed at $t=0$, the water flows downward and outward under gravity.  The problem
is to find the subsequent flow and determine the shape of the
free surface.  This question is addressed in the context of shallow-water
theory, e.g., by Acheson \cite{Ach1990}, and thus serves as a typical hydrodynamic
problem of relevance for CH2 solutions with the $+$ sign choice in (\ref{LL1}).

\subsection*{Plan of the paper}

Section \ref{SpectralProblem-sec} discusses the isospectral problem for the CH2 system.
Section \ref{sec:3} treats asymptotics of the Jost solutions for the CH2 scattering problem.
Section \ref{RHP-sec} explains how analytic solutions for CH2 are obtained by formulating the Inverse Scattering Transform (IST) for CH2 as a Riemann-Hilbert problem (RHP). Perhaps not unexpectedly from the viewpoint of CH2 as a fluids system, its solutions possess the parameterised form (\ref{eq36}-\ref{eq36a}) corresponding to fluid continuum flow. 
Section \ref{solitons} treats multi - soliton solutions of CH2 arising as reflectionless potentials. That is, the reflection coefficient in the inverse scattering transform is taken to vanish in the solution of the RHP for CH2.
Section \ref{EPDiff-CH-eqn-sec} provides a slightly modified CH2 equation that admits peakon solutions, but may not be integrable. 
Section \ref{conclusion-sec} closes the paper by giving a brief summary of its main points and indicating some directions for future research. 

\section{The scattering problem for CH2}\label{SpectralProblem-sec}

\paragraph{Outlook for the CH2 scattering problem.} This section begins our discussion of the isospectral problem for the two-component CH equation with a single velocity, denoted CH2 for simplicity.  The next three short sections will be devoted to further discussions of the CH2 scattering problem. In \S\ref{sec:3}, we will treat the asymptotics of the Jost solutions for the CH2 scattering problem. In \S\ref{RHP-sec}, we will explain how to formulate the Inverse Scattering Transform for CH2 as a Riemann-Hilbert problem. Finally, in \S\ref{solitons} we will derive multi - soliton solutions of CH2 that arise as reflectionless potentials. 

\subsection{CH2 spectral problem}

The {\bfi spectral problem for CH2} (\ref{L1}) is a type of Schr\"odinger equation with an `energy dependent' potential. In particular, it is quadratic in the spectral parameter and the potential functions multiply the spectral parameter. (This is the so-called weighted problem.) It shares some features in common with
Sturm-Liouville spectral problems, see for example \cite{JJ72,K75,SS97}.
An `energy dependent' spectral problem also appears in the inverse scattering transform 
of an integrable generalization of the Bousinesq equation (Kaup-Bousinesq equation)
\cite{K75}.

 Asymptotically, as $|x|\rightarrow \infty$, the spectral problem (\ref{L1}) for CH2 reduces to

\b \label{SPCH2_largex} \Psi_{xx}&=&\Big(-\rho_0^2 \lambda^2
+\omega \lambda +\frac{1}{4}\Big)\Psi,  \e

or, simply 
\b \label{SPCH2_large x1} \Psi_{xx}&=&-k^2\Psi, 
\e
where we introduce a spectral parameter $k$ via the equation

\b \label{k}  -\rho_0^2 \lambda^2+ \omega
\lambda+\frac{1}{4}+k^2=0, \e 
The solutions of (\ref{SPCH2_large x1}) oscillate for real $k$. Consequently, the
continuous spectrum is the real line in the complex $k$-plane.

The quadratic equation (\ref{k}) has roots, 
\b \label{lambda} 
\lambda(k,\sigma)=
\frac{\omega}{2\rho_0^2}+\frac{\sigma k}{\rho_0}\sqrt{1+\frac{\rho_0^2+\omega^2}{4\rho_0^2
k^2}},\e

where $\sigma=\pm 1$, and we assume $\sqrt{w}=\sqrt{|w|}e^{\frac{1}{2}Arg(w)}$
where $0\le Arg(w)<2\pi$. An expansion of (\ref{lambda}) for large $|k|$ yields,
\b \label{lambda_large k}
\lambda(k,\sigma)=\frac{\sigma
k}{\rho_0}+\frac{\omega}{2\rho_0^2}+\frac{\sigma(\rho_0^2+\omega^2)}
{8\rho_0^3}\frac{1}{k}+O\Big(\frac{1}{k^3}\Big).
\e
This expansion uniquely determines $\lambda$ from
$k$ and $\sigma$. Equation (\ref{lambda}) possesses a reflection property that we will assume explicitly from here on, that
\b \label{lambda_minus_k}
\lambda(-k,-\sigma)=\lambda(k,\sigma), \e and also, for real $k$, \
\b \label{lambda_prop}\bar{
\lambda}(k,\sigma)=\lambda(k,\sigma), \e
where $\bar{\lambda}$ is the complex conjugate.

As usual, for real $k\neq 0$ a basis in the space of solutions of (\ref{L1}) can be introduced, fixed by its
asymptotic behavior when $x\rightarrow\infty$  \cite{ZMNP,C01,CGI}: \b \label{eq5} \psi_{1}(x,k)&=&e^{-ikx}+o(1), \qquad
x\rightarrow\infty;
 \\\label{eq6}
\psi_{2}(x,k)&=& e^{ikx}+o(1), \qquad x\rightarrow \infty. \e

A complementary basis can also be introduced, fixed by its asymptotic behavior when
$x\rightarrow -\infty$:

\b \label{eq5a} \varphi_{1}(x,k)&=&e^{-ikx}+o(1), \qquad
x\rightarrow -\infty;
 \\\label{eq6a}
\varphi_{2}(x,k)&=& e^{ikx}+o(1), \qquad x\rightarrow -\infty. \e

\subsection{The scattering matrix, the Jost solutions and the reflection coefficient}
Since $\lambda$ depends not only on $k$ but also on $\sigma$,
it follows that the entire spectral problem, as well as the eigenfunctions,
are labelled by $\sigma$.  For all real $k\neq 0$, if
$\Psi(x,k,\sigma)$ is a solution of (\ref{L1}), then
$\Psi(x,-k,-\sigma)$ is also a solution, since they share the same $\lambda$,
according to (\ref{lambda_minus_k}). Thus, \b \label{eq5aa} \varphi_{1}(x,k,\sigma)=\varphi_{2}(x,
-k,-\sigma), \qquad \psi_{1}(x,k,\sigma)=\psi_{2}(x, -k,-\sigma).
 \e

Due to the reality of $q$, $\rho$ in (\ref{L1}) and the
property (\ref{lambda_prop}) for $\lambda$, for real $k$ we have

\b \label{eq6aa} \varphi_{1}(x,k,\sigma)=\bar{\varphi}_{2}(x,
k,\sigma), \qquad \psi_{1}(x,k,\sigma)=\bar{\psi}_{2}(x,
k,\sigma). \e

For real $k$ the vectors of each of the bases may be represented as a linear
combination of the vectors of the other basis: \b \label{eq7}
\varphi_{i}(x,k,\sigma)=\sum_{l=1,2}T_{il}(k,\sigma)\psi_{l}(x,k,\sigma)
,\e where the matrix  $T(k,\sigma)$ defined above is called the
{\bfi scattering matrix}.

For real $k\neq 0$, instead of $\varphi_{1}(x,k,\sigma)$ and
$\varphi_{2}(x,k,\sigma)$, due to (\ref{eq6aa}), for simplicity we
can write correspondingly $\varphi(x,k,\sigma)$,
$\bar{\varphi}(x,k,\sigma)$. Similarly, we can replace
$\psi_{1,2}(x,k,\sigma)$ by $\psi(x,k,\sigma)$ and its complex conjugate 
$\bar{\psi}(x,k,\sigma)$. Thus $T(k,\sigma)$ has the form
(with real $k$)

\b \label{T} T(k,\sigma) = \left( \begin{array}{cc}
   a(k,\sigma)&  b(k,\sigma)  \\
  \bar{b}(k,\sigma) &  \bar{a}(k,\sigma) \\
\end{array}  \right) \,
\e

and clearly \b \label{eq8}
\varphi(x,k,\sigma)=a(k,\sigma)\psi(x,k,\sigma)+b(k,\sigma)\bar{\psi}(x,k,\sigma).
\e

{\bf Remark.} The solutions $\varphi(x,k,\sigma)$ and $\psi(x,k,\sigma)$ are called the {\bfi Jost solutions}.

The Wronskian $W(f_{1},f_{2})\equiv
f_{1}\partial_{x}f_{2}-f_{2}\partial_{x}f_{1}$ of any pair of
solutions of (\ref{L1}) does not depend on $x$. Therefore, perhaps not unexpectedly,

\b \label{eq9} W(\varphi(x,k,\sigma), \bar{\varphi}(x,k,\sigma))=
W(\psi(x,k,\sigma), \bar{\psi}(x,k,\sigma))=2ik. \e

From (\ref{eq8}) and (\ref{eq9}) it follows that

\b \label{eq10}
|a(k,\sigma)|^2-|b(k,\sigma)|^2=1. \e

Hence, $\det (T(k,\sigma))=1$. That is, the determinant of the scattering matrix is unity.

In analogy with the spectral problem for the KdV equation, (which
is the Schr\"odinger equation from quantum mechanics) \cite{ZMNP},
one can introduce a {\bfi reflection coefficient},
\b \label{reflect-def}
\mathcal{R}(k,\sigma)=b(k,\sigma)/a(k,\sigma)
.\e

%From (\ref{eq10}) it follows that the scattering is unitary, i.e.

%\b \label{eq13} |\mathcal{T}(k)|^{2}+|\mathcal{R}(k)|^{2}=1. \e

The matrix $T(k,\sigma)$ in  (\ref{T}) is determined from the knowledge 
of $\mathcal{R}(k,\sigma)$ for real $k>0$ only. Indeed, from
(\ref{eq5aa}) we have $\psi(x,k,\sigma)=\bar{\psi}(x,-k,-\sigma)$,
etc. for real $k$. Hence, the scattering data satisfy
\b
\bar{a}(k,\sigma)=a(-k,-\sigma)
,\qquad
\bar{b}(k,\sigma)=b(-k,-\sigma)
.
\e
Also, it is sufficient to know
$\mathcal{R}(k,\sigma)$ only on the half line $k>0$, since
$\mathcal{R}(-k,-\sigma)=\bar{\mathcal{R}}(k,\sigma)$.

\section{Asymptotics of the Jost solutions for CH2 as $|k|\to \infty$}\label{sec:3}

\paragraph{Outlook.}
This section continues the analysis of the CH2 scattering problem, by discussing the asymptotic behavior of the Jost solutions.

\subsection{Analyticity properties}
The analyticity properties of the Jost solutions and of $a(k,\sigma)$
play an important role in our considerations. We will need also
the asymptotic behavior of the Jost solutions for $|k|\rightarrow \infty$
which have the form (cf. \cite{CGI,CI06})

\b \psi(x,k,\sigma)= e^{- i k
x-ik\int_{\infty}^x(\frac{\rho(x')}{\rho_0}
-1)dx'+i\frac{\sigma}{2}\int_{\infty}^x(\frac{q(x')}{\rho(x')}
-\frac{\omega \rho(x')}{\rho_0^2})dx'} 
\left[X_0(x)+O\left(\frac{1}{k}\right)\right],\nonumber
\\ \label{psi} \e

\b \varphi(x,k,\sigma)= e^{- i k
x-i k\int_{-\infty}^x(\frac{\rho(x')}{\rho_0}
-1)dx'+i\frac{\sigma}{2}\int_{-\infty}^x(\frac{q(x')}{\rho(x')}
-\frac{\omega \rho(x')}{\rho_0^2})dx'} 
\left[X_0(x)+O\left(\frac{1}{k}\right)\right],  \nonumber
\\   \label{fi}\e

where $X_0(x)=\left(\frac{\rho_0}{\rho(x)}\right)^{1/2}$.
(If initially $\rho(x,0)>0$, one can easily prove that  $\rho(x,t)$ always remains positive.) 
As in \cite{CGI} one can show that $\psi(x,k,\sigma)$ is analytic
in the lower complex half $k$ -plane, while $\varphi(x,k,\sigma)$ is
analytic in the upper complex half $k$ -plane.

The expression for $a(k,\sigma)$ can be extended into the upper half
plane by
\b \label{auhp} a(k,\sigma)=\frac{1}{2ik}
W(\varphi(x,k,\sigma),\psi(x,-k,-\sigma) ). \e
An immediate consequence of (\ref{psi}), (\ref{fi}) and
(\ref{auhp}) is:

\begin{equation}\label{eq:a-as}
    \lim_{k\to \infty} a(k,\sigma) e^{ik\alpha-i\sigma\beta} = 1, \qquad k\in \bbbc_+,
\end{equation}
where the quantities
\b\label{int1} \alpha&=& \int
_{-\infty}^{\infty}\left(\frac{\rho(x)}
{\rho_0}-1\right)\text{d}x ,\\
\beta&=&\frac{1}{2}\int_{-\infty}^{\infty}\left(\frac{q(x)}{\rho(x)}
-\frac{\omega \rho(x)}{\rho_0^2}\right)dx  ,
\label{int2}\e 
are two integrals of the system, \cite{I06}.

\subsection{The discrete spectrum.} 
 
The discrete spectrum can be found as follows. Suppose that
$k_0(\sigma)\in \bbbc_+$ is a zero of $a(k,\sigma)$. Then
$\varphi(x,k_0,\sigma)$ and $\psi(x,-k_0,-\sigma)$ are
linearly dependent (\ref{eq8}): \begin{equation} \label{eq200} \varphi(x,k_0,\sigma)=b_0(\sigma)\psi(x,-k_0,-\sigma).
\end{equation}

From here we see that  $\varphi(x,k_0,\sigma)$
decays exponentially for both $x\rightarrow-\infty$ (which follows
from the definition of $\varphi(x,k_0,\sigma)$) and $x\rightarrow
+ \infty$ (since $\psi(x,-k_0,-\sigma)=e^{ikx}$ for
$x\rightarrow \infty$). Therefore  $\varphi(x,k_0,\sigma)$ is a well
defined eigenfunction of the discrete spectrum with an eigenvalue
$k_0$.

 Now, multiplying (\ref{L1}) by
$\bar{\varphi}(x,k_0,\sigma)$ and performing some manipulations
while keeping in mind that the eigenfunction decays exponentially for
both $x\rightarrow\pm\infty$, we obtain

\b \lambda^2(k_0)\int
_{-\infty}^{\infty}q_2(x)|\varphi|^2\text{d}x+\lambda(k_0)\int
_{-\infty}^{\infty}q_1(x)|\varphi|^2\text{d}x+\nonumber  \\\int
_{-\infty}^{\infty}\Big(\frac{1}{4}|\varphi|^2+|\varphi_x|^2\Big)\text{d}x=0\label{identityforlambda}\e

This identity can be regarded as an equation for $\lambda(k_0)$
where $k_0$ is a parameter. From the quadratic formula,
the two roots $\lambda(k_0,\sigma)$ and $\lambda(k_0,-\sigma)$  satisfy

\b \lambda (k_0,\sigma)\lambda (k_0,-\sigma)=-\,\frac{\int
_{-\infty}^{\infty}\Big(\frac{1}{4}|\varphi|^2+|\varphi_x|^2\Big)\text{d}x}{\int
_{-\infty}^{\infty}\rho^2(x)|\varphi|^2\text{d}x}
\,. \label{lambda12}
\e
On the other hand, from (\ref{k}) we have

\b
\lambda (k_0,\sigma)\lambda (k_0,-\sigma)=-\frac{1}{\rho_0^2}\Big(k_0^2+\frac{1}{4}\Big).
\label{lambda12k}\e

From (\ref{lambda12}) and (\ref{lambda12k}) it follows that
$k_0^2+\frac{1}{4}$ is real and positive. Since $k_0$ is in the
upper half complex plane, it should be exactly on the imaginary
axis, $k_0=i\kappa_0$, where $\kappa_0$ is real, and
$0<\kappa_0<1/2$. With this restriction on $k_0$, notice that 
$\lambda (k_0,\sigma)$ is real.

Let us show that $a(k,\sigma)$ can have only simple zeroes in the
upper half complex plane.  The dot will be used to denote the
derivatives with respect to $k$ at the point $k_0$. From (\ref{k})
we have $\dot{\lambda}=2k/(2\rho_0^2 \lambda-\omega)$.
Differentiating the (\ref{L1}) (written for the eigenfunction
$\varphi$) with respect to $k$ and multiplying by $\bar{\varphi}$ we
obtain

\b (\bar{\varphi}\dot{\varphi}_x-\bar{\varphi}_x\dot{\varphi})|_{-\infty}^{\infty}=
\dot{\lambda}\int _{-\infty}^{\infty}(q-2\lambda \rho^2)|\varphi|^2\text{d}x. \label{fidot1} \e

Next, using the asymptotics
$\dot{\varphi}(x,k_0,\sigma)\rightarrow
\dot{a}(k_0,\sigma)e^{\kappa_0 x}$,
$\varphi(x,k_0,\sigma)\rightarrow b_0(\sigma) e^{-\kappa_0 x}$ for
$x\rightarrow\infty$; $\dot{\varphi}(x,k_0,\sigma)\rightarrow 0$,
$\varphi(x,k_0,\sigma)\rightarrow 0$ for $x\rightarrow-\infty$,
(\ref{fidot1}) can be transformed into

\b \int _{-\infty}^{\infty}(q-2\lambda(k_0)
\rho^2)|\varphi|^2\text{d}x=(\omega-2\rho_0^2 \lambda(k_0))i \bar{b}_0
\dot{a}(k_0). \label{fidot2} \e

As a corollary we notice that the quantity $R_0(\sigma,t)=\frac{b_0}{i\dot{a}(k_0)}$
is real.

If $\int _{-\infty}^{\infty}(q-2\lambda(k_0)
\rho^2)|\varphi|^2\text{d}x\neq 0$, then $\dot{a}(k_0)\neq 0$ and
the zero $k_0$ is simple. Therefore, a multiple zero is possible,
only if \b \int _{-\infty}^{\infty}(q-2\lambda(k_0)
\rho^2)|\varphi|^2\text{d}x= 0. \label{int0} \e

Suppose that (\ref{int0}) is satisfied. From (\ref{L1})
(written for the eigenfunction $\varphi$) multiplied by $\bar{\varphi}$
we obtain

\b 
-\lambda^2(k_0)\int
_{-\infty}^{\infty}\rho^2(x)|\varphi|^2\text{d}x
&+&
\lambda(k_0)\int
_{-\infty}^{\infty}q(x)|\varphi|^2\text{d}x 
\nonumber  \\
&&
+ \int
_{-\infty}^{\infty}\Big(\frac{1}{4}|\varphi|^2+|\varphi_x|^2\Big)\text{d}x=0
.\label{identityforlambdak0}
\e

From (\ref{int0}) and (\ref{identityforlambdak0}) we obtain

\b \lambda(k_0)^2=-\frac{\int
_{-\infty}^{\infty}\Big(\frac{1}{4}|\varphi|^2+|\varphi_x|^2\Big)\text{d}x}{\int
_{-\infty}^{\infty}\rho^2(x)|\varphi|^2\text{d}x}, \label{lambdak0} \e

but from $(\ref{identityforlambdak0})$ itself we find that the
product of the two roots is

\b \lambda(k_0, \sigma)\lambda(k_0, - \sigma)=-\frac{\int
_{-\infty}^{\infty}\Big(\frac{1}{4}|\varphi|^2+|\varphi_x|^2\Big)\text{d}x}{\int
_{-\infty}^{\infty}\rho^2(x)|\varphi|^2\text{d}x}. \label{lambdak012}\e

From (\ref{lambdak0}), (\ref{lambdak012}) and (\ref{lambda}) we
find that
$\lambda(k_0,\sigma)=\lambda(k_0,-\sigma)=\frac{\omega}{2\rho_0^2}$. Thus
in this case the multiplier $\omega-2\rho_0^2 \lambda(k_0)$ on
the right hand side of (\ref{fidot2}) is also zero. Therefore, we
can use l'Hospital's rule in the evaluation of $\dot{a}(k_0)$.
From (\ref{fidot2})  we have \b \dot{a}(k_0)&=&\lim_{\lambda\rightarrow
\frac{\omega}{2\rho_0^2}}\frac{\frac{\partial}{\partial
\lambda}\int _{-\infty}^{\infty}(q-2\lambda
\rho^2)|\varphi|^2\text{d}x}{i \bar{b}_0\frac{\partial}{\partial
\lambda}(\omega-2 \lambda \rho_0^2) } \nonumber\\
&=&\lim_{\lambda\rightarrow
\frac{\omega}{2\rho_0^2}}\frac{-2\int
_{-\infty}^{\infty}\rho^2|\varphi|^2\text{d}x+2\frac{\partial
k}{\partial \lambda}\int _{-\infty}^{\infty}(q-2\lambda \rho^2)|\varphi||
\dot{\varphi}| \text{d}x}{-2i \bar{b}_0 \rho_0^2} \nonumber \\
&=&\frac{\int _{-\infty}^{\infty}\rho^2|\varphi|^2\text{d}x}{i \bar{b}_0
\rho_0^2}, \label{adot} \e

since \b \lim_{\lambda\rightarrow
\frac{\omega}{2\rho_0^2}}\frac{\partial k}{\partial
\lambda}=\lim_{\lambda\rightarrow
\frac{\omega}{2\rho_0^2}}\frac{2\rho_0^2
\lambda-\omega}{2k}=0. \nonumber \e

 Then (\ref{adot}) shows
that $\dot{a}(k_0)\ne 0$, i.e. $k_0$ is a simple zero of $a(k)$.

\subsection{Summary of asymptotic behavior of Jost functions for CH2}
 To summarize: the discrete spectrum in the upper half plane
consists of finitely many points $k_{n}=i\kappa
_{n}$, $n=1,\ldots,N$, which are the simple zeroes of
$a(k,\sigma)$. Furthermore, each $\kappa_{n}$ is real and $0<\kappa_n<1/2$. 

\paragraph{Eigenfunctions.}
Two eigenfunctions $\varphi^{(n)}(x,\sigma)$ belong to 
each eigenvalue $i\kappa_n$, because there are two eigenvalues 
$\lambda_n(\sigma)=\lambda(i\kappa_n, \sigma)$ that correspond to a given $\kappa_n$. We can take this
eigenfunction to be \b \label{eq201}\varphi^{(n)}(x, \sigma)\equiv
\varphi(x,i\kappa_n,\sigma)
\,.\e

The asymptotic behavior of $\varphi^{(n)}$, according to (\ref{eq5a}),
(\ref{eq6}), (\ref{eq200}) is

\b \label{eq203} 
\varphi^{(n)}(x,\sigma)
&=&
e^{\kappa_n x}
+ o(e^{\kappa_n x})
 \quad \hbox{for} \quad
x\rightarrow  - \infty
\,,
 \\\label{eq204}
\varphi^{(n)}(x,\sigma)
&=& 
b_n(\sigma) e^{-\kappa_n x}
+
o(e^{-\kappa_n x})
\quad \hbox{for} \quad
x\rightarrow \infty. \e

%The sign of $b_n$ obviously depends on the number of the
%zeroes of $\varphi^{(n)}$. Suppose that

%\b \label{eqN} 0<\kappa_{1}<\kappa_{2}<\ldots<\kappa_{N}<1/2.\e
%Then from the oscillation theorem for the Sturm-Liouville problem
%\cite{B}, $\varphi^{(n)}$ has exactly $n-1$ zeroes. Therefore

%\b \label{eq205} b_n= (-1)^{n-1}|b_n|.\e

\paragraph{Scattering data.}
The set \b \label{eq206} \mathcal{S}\equiv\{ \mathcal{R}(k,\sigma)\quad
(k>0),\quad \kappa_n,\quad b_n(\sigma),\quad n=1,\ldots N,
\quad \sigma=\pm 1 \},  \e

is called the {\bfi scattering data}.

The time evolution of the scattering data can be easily obtained as
follows.

The second equation of the Lax pair is \b
\Psi_{t}&=&\Big(\frac{1}{2\lambda}-u\Big)\Psi_{x}+\frac{u_{x}}{2}\Psi+\gamma\Psi,
\label{Laxt} \e

where we introduced an arbitrary constant $\gamma$ (which does not affect
the compatibility).

From (\ref{eq8}) with $x\rightarrow\infty$ one has \b \label{eq14} \varphi(x,k,\sigma)=a(k,
\sigma)e^{-ikx}+b(k,\sigma)e^{ikx}+o(1). \e The substitution of
$\varphi(x,k,\sigma)$ into (\ref{Laxt}) with $x\rightarrow\infty$
gives

\b \label{eq15} \varphi_{t}=\frac{1}{2\lambda}\varphi_{x}+\gamma
\varphi 
\,.\e

From (\ref{eq14}), (\ref{eq15}) with the choice
$\gamma=ik/2\lambda$ for the eigenfunction $\varphi(x,k)$ we obtain

\b \label{eq16} a_t(k,\sigma,t)&=&0,
 \\\label{eq17}
b_t(k,t,\sigma)&=& \frac{i k}{\lambda }b(k,t,\sigma). \e

Thus, we find 
 \b \label{eq18} a(k,t,\sigma)=a(k,0,\sigma), \qquad
b(k,t,\sigma)=b(k,0,\sigma)e^{\frac{i k}{\lambda }t}, \e \b \label{eq19}
\mathcal{R}(k,t,\sigma)=\mathcal{R}(k,0,\sigma)e^{\frac{i
k}{\lambda(k,\sigma) }t}. \e

In other words, $a(k,\sigma)$ does not depend on $t$ and can serve
as a generating function of the conservation laws.

\paragraph{Time evolution of the data on the discrete spectrum.}
The time evolution of the data on the discrete spectrum is found
as follows. Let us introduce the notation
$\dot{a}_n(\sigma)\equiv \dot{a}(i\kappa_n,\sigma)$. We
notice that $i\kappa_n$ are zeroes of $a(k,\sigma)$, which
does not depend on $t$, and hence $(\kappa_n)_t=0$ and
$\lambda_n(\sigma)_t=0$. From (\ref{Laxt}), (\ref{lambda}) with
$\gamma=ik/2\lambda$; $k=i\kappa_n$ and (\ref{eq204}) one
can obtain
 \b \label{eq207} b_n(\sigma)_t=\frac{-\kappa_n
}{\lambda_n(\sigma)}b_n(\sigma). \e

It is convenient to use the variable $R_n(\sigma)\equiv
\frac{b_n(\sigma)}{i\dot{a}_n(\sigma)}$, which is real, according to (\ref{fidot2})
and evolves with $t$ as \b \label{eq207}
R_n(t,\sigma)=R_n(0,\sigma)\exp\Big(\frac{-\kappa_n
}{\lambda_n(\sigma)}t\Big). \e

When $k$ is in the upper half plane one can derive the following dispersion
relation for $a(k, \sigma)$ e.g. following the pattern for the CH case from \cite{CI06}:

\begin{equation}\label{eqi21}
 \ln a(k,\sigma)=-i\alpha k +i\sigma \beta+\sum
 _{n=1}^{N}\ln\frac{k-i\kappa_n}{k+i\kappa_n}+
 \frac{1}{\pi i}\int _{-\infty}^{\infty}\frac{\ln |a(k',\sigma)|}{k'-k}dk',
\end{equation}
i.e. $a(k,\sigma)$ is determined by $|a(k,\sigma)|$ given on the real
line $k\in \mathbb{R}$.

\paragraph{Outlook.}
In the next section, we will develop the Inverse Scattering
Transform for CH2.  The special case of reflectionless potentials 
($\mathcal{R}(k,\sigma)=0$ for all $k$)
corresponds to an important class of solutions, namely the
multi-soliton solutions. These will be separately studied in Section \ref{solitons}, where a 
formula for the $N$-soliton solution will be obtained.

\section{Analytic solutions and the Riemann-Hilbert Problem for CH2}\label{RHP-sec}

This section explains how analytic solutions for CH2 are obtained by formulating the Inverse Scattering Transform for CH2 as a Riemann-Hilbert problem (RHP).

\subsection{Preliminaries}

We begin by introducing the following new variables, cf. the integrals of motion in equation (\ref{int1}) and (\ref{int2}) 
\b\label{eq22} y(x)&=&x+\int_{\infty}^x\left(\frac{\rho(x')}{\rho_0}-1\right)\text{d}x',\\
z(x)&=&\frac{1}{2}\int_{\infty}^{x}\left(\frac{q(x')}{\rho(x')}
-\frac{\omega \rho(x')}{\rho_0^2}\right)\text{d}x'.
\label{eq22a}\e

In terms of these variables, the expansion (\ref{psi}) may be written as

\b\label{eq23} \psi(x,k,\sigma)=e^{-iky+i\sigma z} \left[X_0(x)+O\left(\frac{1}{k}\right)\right].\e

Furthermore, the function $\underline{\chi}(x,k,\sigma)\equiv
\psi(x,k,\sigma)e^{ikx}$ is analytic for $\mathrm{Im}\phantom{*}
k<0$, due to arguments similar to these, given in \cite{C01} for
the CH case. This follows from the representation

\b\label{eq24}
\underline{\chi}(x,k)=1-\int_{x}^{\infty}\frac{e^{2ik(x-x')}-1}{2ik}[-\lambda^2
(\rho^2(x')-\rho_0^2)+\lambda(q(x')-\omega))]\,\underline{\chi}(x',k)dx'.\nonumber
\e

Notice that $y(x)$ is a bounded function for all $x$, which follows from the assumption that $\rho(x)-\rho_0$ is a Schwartz class function. Therefore, the function 
\b \label{eq25}\underline{\psi}(x,k,\sigma)\equiv
\psi(x,k,\sigma)e^{iky-i\sigma z}= X_0(x)+O\left(\frac{1}{k}\right)\e 
is also analytic for $\mathrm{Im}\phantom{*} k<0$.

Similarly, \b \label{eq26} \underline{\varphi}(x,k,\sigma) &\equiv&
\varphi(x,k,\sigma)e^{i k y-i\sigma z +i k \alpha-i\sigma \beta}
=X_0(x)+O\left(\frac{1}{k}\right)\e

is analytic for $\mathrm{Im}\phantom{*} k>0$.

Multiplying (\ref{eq8}) by $e^{i k y - i\sigma z}/a(k,\sigma)$ and
using (\ref{eq25}) and (\ref{eq26}) we obtain

\b \label{eq27}
\frac{\underline{\varphi}(x,k,\sigma)}{e^{ik\alpha-i\sigma\beta}a(k,\sigma)}=
\underline{\psi}(x,k,\sigma)+\mathcal{R}(k,\sigma)\underline{\psi}(x,-k,-\sigma)e^{2
i k y-2i\sigma z}.
\e

The function
$\frac{\underline{\varphi}(x,k,\sigma)}{e^{ik\alpha-i\sigma\beta}a(k,\sigma)}$
is analytic for $\mathrm{Im}\phantom{*} k>0$, while 
$\underline{\psi}(x,k,\sigma)$ is analytic for
$\mathrm{Im}\phantom{*} k<0$. Thus, equation (\ref{eq27}) represents an
additive Riemann-Hilbert Problem (RHP) with a jump on the real line,
given by
$\mathcal{R}(k,\sigma)\underline{\psi}(x,-k,-\sigma)e^{2i k y - 2 i \sigma
z}$ and a normalization condition, $\lim_{|k|\to \infty}\underline{\psi}(x,k,\sigma)=X_0(x)$.

\subsection{Solving the Riemann-Hilbert Problem for CH2}
In this section we will follow the standard technique for solving RHP. We
integrate the two analytic functions with respect to $\oint \frac{d k'}{k'-k} (\cdot)$ over the boundary of their analyticity domains, using the normalization
condition. In our case the domains (the upper $\mathbb{C}_{+}$ and the lower
$\mathbb{C}_{-}$ complex half-planes) have
the real line as a common boundary and there we relate the integrals using
the jump condition. The RHP approach for the CH equation is presented in
\cite{CGI,dMonv, HI10}, for the Kaup-Bousinesq equation (which also has energy-dependent
spectal problem) in \cite{SS97}.

Let us take an arbitrary $k$ from the lower half plane
($\mathrm{Im}\phantom{*} k<0$). Then using the Residue Theorem,
(\ref{int1}) and (\ref{eq200}) we can compute the integral \b
\label{eq30}I&=& \frac{1}{2\pi
i}\oint_{C_+}\frac{\underline{\varphi}(x,k',\sigma)}{e^{ik'\alpha-i\sigma\beta}
a(k',\sigma)}\frac{d k'}{k'-k}\nonumber\\
&=& \sum_{n=1} ^{N}
\frac{\underline{\varphi}(x,i\kappa_n,\sigma)}{(i\kappa_n-k)e^{-\kappa_n\alpha
-i\sigma\beta}\dot{a}_n(\sigma)}\nonumber\\
&=&\sum_{n=1} ^{N}
\frac{i R_n(\sigma)e^{-2\kappa_n y-2i \sigma z}\underline{\psi}
(x,-i\kappa_n,-\sigma)}{i\kappa_n-k},\e

where $C_+$ is the closed contour in the upper half plane
(Fig. \ref{fig:1}).

\begin{center}

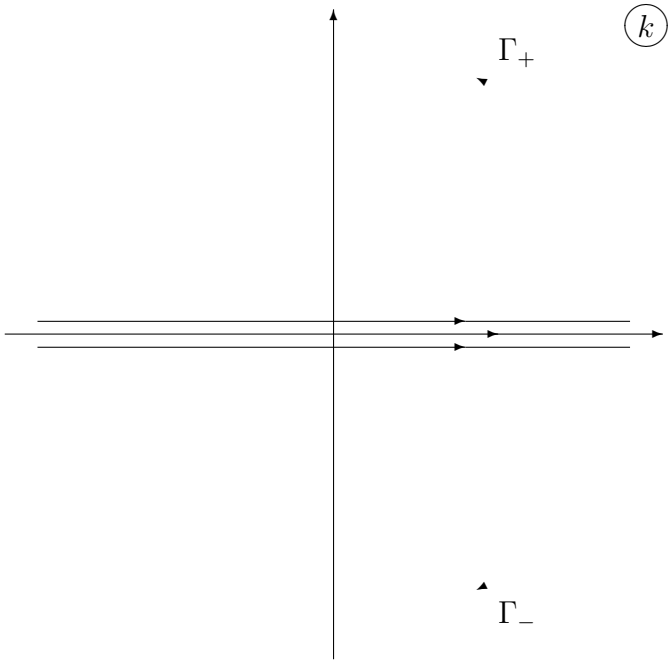
\begin{figure}
%\input{b03-1.mtx}
% GNUPLOT: LaTeX picture with emtex specials
%\setlength{\unitlength}{0.240900pt}
\setlength{\unitlength}{0.160900pt}
\ifx\plotpoint\undefined\newsavebox{\plotpoint}\fi
\sbox{\plotpoint}{\rule[-0.175pt]{0.350pt}{0.350pt}}%
\special{em:linewidth 0.7pt}%
\begin{picture}(1875,1800)(0,0)
\put(1772,1649){\makebox(0,0)[c]{$k$}} \put(1772,1649){\circle{100}}
\put(1424,1585){\makebox(0,0)[l]{$\Gamma_{+}$}}
\put(1424,260){\makebox(0,0)[l]{$\Gamma_{-}$}}
\put(264,923){\vector(1,0){1547}} \put(341,953){\vector(1,0){1006}}
\put(1347,953){\line(1,0){387}} \put(1347,892){\line(1,0){387}}
\put(341,892){\vector(1,0){1006}} \put(1037,158){\vector(0,1){1529}}
\put(1386,1518){\special{em:moveto}}
\put(1373,1525){\special{em:lineto}}
\put(1373,1525){\vector(-2,1){0}} \put(1409,923){\vector(1,0){15}}
\put(1386,327){\special{em:moveto}}
\put(1373,320){\special{em:lineto}}
\put(1373,320){\vector(-2,-1){0}}
\put(1733,953){\special{em:moveto}}
\put(1732,974){\special{em:lineto}}
\put(1730,995){\special{em:lineto}}
\put(1727,1016){\special{em:lineto}}
\put(1724,1037){\special{em:lineto}}
\put(1720,1058){\special{em:lineto}}
\put(1716,1079){\special{em:lineto}}
\put(1710,1099){\special{em:lineto}}
\put(1704,1120){\special{em:lineto}}
\put(1698,1140){\special{em:lineto}}
\put(1691,1160){\special{em:lineto}}
\put(1683,1180){\special{em:lineto}}
\put(1675,1199){\special{em:lineto}}
\put(1666,1218){\special{em:lineto}}
\put(1656,1238){\special{em:lineto}}
\put(1646,1256){\special{em:lineto}}
\put(1636,1275){\special{em:lineto}}
\put(1624,1293){\special{em:lineto}}
\put(1612,1310){\special{em:lineto}}
\put(1600,1328){\special{em:lineto}}
\put(1587,1345){\special{em:lineto}}
\put(1574,1361){\special{em:lineto}}
\put(1560,1377){\special{em:lineto}}
\put(1545,1393){\special{em:lineto}}
\put(1530,1408){\special{em:lineto}}
\put(1515,1423){\special{em:lineto}}
\put(1499,1437){\special{em:lineto}}
\put(1483,1451){\special{em:lineto}}
\put(1466,1465){\special{em:lineto}}
\put(1449,1477){\special{em:lineto}}
\put(1432,1490){\special{em:lineto}}
\put(1414,1501){\special{em:lineto}}
\put(1395,1513){\special{em:lineto}}
\put(1377,1523){\special{em:lineto}}
\put(1358,1533){\special{em:lineto}}
\put(1339,1543){\special{em:lineto}}
\put(1319,1552){\special{em:lineto}}
\put(1299,1560){\special{em:lineto}}
\put(1279,1568){\special{em:lineto}}
\put(1259,1575){\special{em:lineto}}
\put(1239,1581){\special{em:lineto}}
\put(1218,1587){\special{em:lineto}}
\put(1197,1592){\special{em:lineto}}
\put(1176,1597){\special{em:lineto}}
\put(1155,1601){\special{em:lineto}}
\put(1134,1604){\special{em:lineto}}
\put(1113,1607){\special{em:lineto}}
\put(1091,1609){\special{em:lineto}}
\put(1070,1610){\special{em:lineto}}
\put(1048,1610){\special{em:lineto}}
\put(1027,1610){\special{em:lineto}}
\put(1005,1610){\special{em:lineto}}
\put(984,1609){\special{em:lineto}}
\put(962,1607){\special{em:lineto}}
\put(941,1604){\special{em:lineto}}
\put(920,1601){\special{em:lineto}}
\put(899,1597){\special{em:lineto}}
\put(878,1592){\special{em:lineto}}
\put(857,1587){\special{em:lineto}}
\put(836,1581){\special{em:lineto}}
\put(816,1575){\special{em:lineto}}
\put(796,1568){\special{em:lineto}}
\put(776,1560){\special{em:lineto}}
\put(756,1552){\special{em:lineto}}
\put(736,1543){\special{em:lineto}}
\put(717,1533){\special{em:lineto}}
\put(698,1523){\special{em:lineto}}
\put(680,1513){\special{em:lineto}}
\put(661,1501){\special{em:lineto}}
\put(643,1490){\special{em:lineto}}
\put(626,1477){\special{em:lineto}}
\put(609,1465){\special{em:lineto}}
\put(592,1451){\special{em:lineto}}
\put(576,1437){\special{em:lineto}}
\put(560,1423){\special{em:lineto}}
\put(545,1408){\special{em:lineto}}
\put(530,1393){\special{em:lineto}}
\put(515,1377){\special{em:lineto}}
\put(501,1361){\special{em:lineto}}
\put(488,1345){\special{em:lineto}}
\put(475,1328){\special{em:lineto}}
\put(463,1310){\special{em:lineto}}
\put(451,1293){\special{em:lineto}}
\put(439,1275){\special{em:lineto}}
\put(429,1256){\special{em:lineto}}
\put(419,1238){\special{em:lineto}}
\put(409,1218){\special{em:lineto}}
\put(400,1199){\special{em:lineto}}
\put(392,1180){\special{em:lineto}}
\put(384,1160){\special{em:lineto}}
\put(377,1140){\special{em:lineto}}
\put(371,1120){\special{em:lineto}}
\put(365,1099){\special{em:lineto}}
\put(359,1079){\special{em:lineto}}
\put(355,1058){\special{em:lineto}}
\put(351,1037){\special{em:lineto}}
\put(348,1016){\special{em:lineto}}
\put(345,995){\special{em:lineto}}
\put(343,974){\special{em:lineto}}
\put(342,953){\special{em:lineto}}
\sbox{\plotpoint}{\rule[-0.350pt]{0.700pt}{0.700pt}}%
\special{em:linewidth 0.7pt}%
\put(342,892){\special{em:moveto}}
\put(343,871){\special{em:lineto}}
\put(345,850){\special{em:lineto}}
\put(348,829){\special{em:lineto}}
\put(351,808){\special{em:lineto}}
\put(355,787){\special{em:lineto}}
\put(359,766){\special{em:lineto}}
\put(365,746){\special{em:lineto}}
\put(371,725){\special{em:lineto}}
\put(377,705){\special{em:lineto}}
\put(384,685){\special{em:lineto}}
\put(392,665){\special{em:lineto}}
\put(400,646){\special{em:lineto}}
\put(409,627){\special{em:lineto}}
\put(419,607){\special{em:lineto}}
\put(429,589){\special{em:lineto}}
\put(439,570){\special{em:lineto}}
\put(451,552){\special{em:lineto}}
\put(463,535){\special{em:lineto}}
\put(475,517){\special{em:lineto}}
\put(488,500){\special{em:lineto}}
\put(501,484){\special{em:lineto}}
\put(515,468){\special{em:lineto}}
\put(530,452){\special{em:lineto}}
\put(545,437){\special{em:lineto}}
\put(560,422){\special{em:lineto}}
\put(576,408){\special{em:lineto}}
\put(592,394){\special{em:lineto}}
\put(609,380){\special{em:lineto}}
\put(626,368){\special{em:lineto}}
\put(643,355){\special{em:lineto}}
\put(661,344){\special{em:lineto}}
\put(680,332){\special{em:lineto}}
\put(698,322){\special{em:lineto}}
\put(717,312){\special{em:lineto}}
\put(736,302){\special{em:lineto}}
\put(756,293){\special{em:lineto}}
\put(776,285){\special{em:lineto}}
\put(796,277){\special{em:lineto}}
\put(816,270){\special{em:lineto}}
\put(836,264){\special{em:lineto}}
\put(857,258){\special{em:lineto}}
\put(878,253){\special{em:lineto}}
\put(899,248){\special{em:lineto}}
\put(920,244){\special{em:lineto}}
\put(941,241){\special{em:lineto}}
\put(962,238){\special{em:lineto}}
\put(984,236){\special{em:lineto}}
\put(1005,235){\special{em:lineto}}
\put(1027,235){\special{em:lineto}}
\put(1048,235){\special{em:lineto}}
\put(1070,235){\special{em:lineto}}
\put(1091,236){\special{em:lineto}}
\put(1113,238){\special{em:lineto}}
\put(1134,241){\special{em:lineto}}
\put(1155,244){\special{em:lineto}}
\put(1176,248){\special{em:lineto}}
\put(1197,253){\special{em:lineto}}
\put(1218,258){\special{em:lineto}}
\put(1239,264){\special{em:lineto}}
\put(1259,270){\special{em:lineto}}
\put(1279,277){\special{em:lineto}}
\put(1299,285){\special{em:lineto}}
\put(1319,293){\special{em:lineto}}
\put(1339,302){\special{em:lineto}}
\put(1358,312){\special{em:lineto}}
\put(1377,322){\special{em:lineto}}
\put(1395,332){\special{em:lineto}}
\put(1414,344){\special{em:lineto}}
\put(1432,355){\special{em:lineto}}
\put(1449,368){\special{em:lineto}}
\put(1466,380){\special{em:lineto}}
\put(1483,394){\special{em:lineto}}
\put(1499,408){\special{em:lineto}}
\put(1515,422){\special{em:lineto}}
\put(1530,437){\special{em:lineto}}
\put(1545,452){\special{em:lineto}}
\put(1560,468){\special{em:lineto}}
\put(1574,484){\special{em:lineto}}
\put(1587,500){\special{em:lineto}}
\put(1600,517){\special{em:lineto}}
\put(1612,535){\special{em:lineto}}
\put(1624,552){\special{em:lineto}}
\put(1636,570){\special{em:lineto}}
\put(1646,589){\special{em:lineto}}
\put(1656,607){\special{em:lineto}}
\put(1666,627){\special{em:lineto}}
\put(1675,646){\special{em:lineto}}
\put(1683,665){\special{em:lineto}}
\put(1691,685){\special{em:lineto}}
\put(1698,705){\special{em:lineto}}
\put(1704,725){\special{em:lineto}}
\put(1710,746){\special{em:lineto}}
\put(1716,766){\special{em:lineto}}
\put(1720,787){\special{em:lineto}}
\put(1724,808){\special{em:lineto}}
\put(1727,829){\special{em:lineto}}
\put(1730,850){\special{em:lineto}}
\put(1732,871){\special{em:lineto}}
\put(1733,892){\special{em:lineto}}
\end{picture}
\caption{The contours $\Gamma _\pm $}
\label{fig:1}
\end{figure}
\end{center}

On the other hand, because of (\ref{eq27}) the same integral can be
computed directly as

\b \label{eq31} I&=& \frac{1}{2\pi
i}\int_{-\infty}^{\infty}\left(\underline{\psi}(x,k',\sigma)+\mathcal{R}(k',\sigma)
\underline{\psi}(x,-k',-\sigma)e^{2i k' y -2 i \sigma z}\right)\frac{d
k'}{k'-k}\nonumber \\&+&\frac{1}{2\pi
i}\int_{\Gamma_+}\frac{\underline{\varphi}(x,k',\sigma)}{e^{i k'\alpha-i\sigma\beta}a(k',\sigma)}\frac{d
k'}{k'-k},\e

where $\Gamma_+$ is the infinite semicircle in the upper half
plane (Fig. \ref{fig:1}). Using the expansion (\ref{eq26}) and limit
(\ref{eq:a-as}), it is straightforward to compute that the
integral over $\Gamma_+$ is simply $(1/2)X_0(x)$.

Similarly,

\b \label{eq32} -\underline{\psi}(x,k,\sigma)&=&\frac{1}{2\pi
i}\oint_{C_-}\underline{\psi}(x,k',\sigma)\frac{d k'}{k'-k}=
\frac{1}{2\pi
i}\int_{-\infty}^{\infty}\underline{\psi}(x,k',\sigma)\frac{d
k'}{k'-k}\nonumber \\&+&\frac{1}{2\pi
i}\int_{\Gamma_-}\underline{\psi}(x,k',\sigma)\frac{d k'}{k'-k},\e
where $C_-$ is the closed contour in the lower half plane,
$\Gamma_-$ is the infinite semicircle in the lower half plane
(Fig. \ref{fig:1}). Due to (\ref{eq23}), (\ref{eq25}) the integral over
$\Gamma_-$ is $-(1/2)X_0(x)$.

Now, from (\ref{eq30}) -- (\ref{eq32}) it follows that for
$\mathrm{Im}\phantom{*} k<0$,

\b \label{eq33}
\underline{\psi}(x,k,\sigma)&=&X_0(x)+
\int_{-\infty}^{\infty}\mathcal{R}(k',\sigma)\underline{\psi}(x,-k',-\sigma)e^{2i
k'y-2i\sigma z}\frac{d k'}{k'-k}\nonumber \\ &+&\sum_{n=1}
^{N}
\frac{iR_n(\sigma)e^{-2\kappa_n y-2i \sigma z}
\underline{\psi}(x,-i\kappa_n,-\sigma)}{k-i\kappa_n}.\e

The expression (\ref{eq33}), taken at $k=-i\kappa_p$,
$p=1,\ldots,N $ gives

\b \label{eq34}
\underline{\psi}(x,-i\kappa_p,\sigma)&=&X_0(x)\nonumber
\\&+&\int_{-\infty}^{\infty}\mathcal{R}(k',\sigma)\underline{\psi}(x,-k',-\sigma)
e^{2i k' y -2i \sigma z}\frac{d k'}{k'+i\kappa_p}\nonumber \\
&-&\sum_{n=1} ^{N}
\frac{R_n(\sigma)e^{-2\kappa_n y - 2i \sigma z}\underline{\psi}(x,-i\kappa_n,-\sigma)}
{\kappa_p+\kappa_n}.\e

From (\ref{eq34}) it follows immediately that

\b
\underline{\psi}(x,-i\kappa_p,-\sigma)&=& X_0(x)\nonumber
\\&+& \int_{-\infty}^{\infty}\mathcal{R}(k',-\sigma)\underline{\psi}(x,-k',\sigma)e^{-2i
k' y+ 2i\sigma z }\frac{d
k'}{k'+i\kappa_p}\nonumber \\
&-&\sum_{n=1} ^{N}
\frac{R_n(-\sigma)e^{-2\kappa_n y + 2 i \sigma z}
\underline{\psi}(x,-i\kappa_n,\sigma)}{\kappa_p+\kappa_n}. \nonumber\\
 \label{eq34a}\e

Equations (\ref{eq33}) -- (\ref{eq34a}) represent a linear
system, from which $\underline{\psi}(x,k,\sigma)$ (for real $k$)
and $\underline{\psi}(x,-i\kappa_n,\pm \sigma)$  can be
expressed through $y$, $z$, which are as yet unknown
functions of $x$ (since $X_0(x)$ can be obtained from $y(x)$).

Finally, we need to find the dependence of $y$ on $x$. From the quadratic roots in 
(\ref{lambda}) we notice that
\[\lambda(-i/2,\sigma)=\frac{\omega - \sigma |\omega|}{2\rho_0^2}.\]
Hence, if we take $\sigma =\sigma_1 \equiv \text{sign}(\omega)$, then 
we have $\lambda(-i/2,\sigma_1)=0$.  Now $\psi(x,k,\sigma_1)$
does not depend on $q$ and $\rho$ for $\lambda=0$ and since
$\psi(x,k,\sigma_1)$ is defined by its asymptotic behavior as $x\to\infty$, which is
$e^{-x/2}$, i.e., it is real when $k=-i/2$. Consequently, we have 
\b \frac{1}{2}\left(\psi(x,-i/2,\sigma_1)+ \text{compl. conj.}\right)=e^{-x/2}. \nonumber \e 

Thus, for
$k=-i/2$, $\sigma =\sigma_1 \equiv \text{sign}(\omega_1)$, equation 
(\ref{eq33}) gives

\b
e^{(-x+y)/2}&=&e^{i\sigma_1 z}\Big(X_0(x)+
\int_{-\infty}^{\infty}\mathcal{R}(k',\sigma_1)\underline{\psi}
(x,-k',-\sigma_1)e^{2i k' y- 2i \sigma_1 z}\frac{d k'}{k'+i/2}\nonumber \\ 
&-& \sum_{n=1}^{N}
\frac{R_n(\sigma_1)e^{-2\kappa_n y - 2i \sigma_1 z}
\underline{\psi}(x,-i\kappa_n,-\sigma_1)}{1/2+\kappa_n}
 \Big)+ \text{complex conjugate}   \label{eq35}\e

In other words,
(\ref{eq33}) -- (\ref{eq35}) represent a system of singular
integral equations for $\underline{\psi}(x,k,\sigma)$ (for real
$k$) and $\underline{\psi}(x,-i\kappa_n(\sigma),\sigma)$,
$\underline{\bar{\psi}}(x,-i\kappa_n(\sigma),\sigma)$.

A similar relation to (\ref{eq33}) can be written if $k\in \mathbb{C}_{+}$ and in particular when $k=i/2$ we have another equation of the type of (\ref{eq35}). Thus, one can recover 
\b
x=X(y,z(y))
\quad\hbox{and}\quad
\frac{dx}{dy}\equiv X_0^2(y,z(y))
.
\e
Then eliminating $z$ eventually yields $x=X(y)$.

% Since $\underline{\psi}(x,k,\sigma)$ (for real $k$) and
%$\underline{\psi}(x,-i\kappa_n(\sigma),\sigma)$,
%$\underline{\bar{\psi}}(x,-i\kappa_n(\sigma),\sigma)$  are known
%from (\ref{eq33}) -- (\ref{eq34a}), $z$ can be expressed
%through $y$ from (\ref{+-sigma}) and finally the equation
%(\ref{eq35}) is a first order differential equation for $y$, since $X_0=\sqrt{X_y}$,
%if $x=X(y) $  represents $x$ via $y$.
%

\paragraph{Parametric forms of the solutions.}
Since the time evolution of the scattering data is known
(\ref{eq207}), the dependence on $t$, i.e. $x=X(y,t)$, is also known, as expressed by the scattering data. Thus the set $\mathcal{S}$ of scattering data (\ref{eq206}) uniquely determines the solution.
From (\ref{eq22}) one obtains the parametric forms of the solutions,
\b \label{eq36}
x&=&X(y,t),\\
\rho(x,t)&=&\frac{\rho_0}{X_y(y,t)},\\
u(x,t)&=&X_t(y,t).
\label{eq36a} \e
These correspond precisely to the relations between Eulerian and Lagrangian variables in compressible ideal fluid dynamics. 

\section{Reflectionless potentials and solitons for CH2}\label{solitons}

This section discusses the simplification of the inverse scattering problem for CH2 in the important case of the
so-called reflectionless potentials, when the scattering data is confined to the case $\mathcal{R}(k,\sigma)=0$ for all real $k$. This class of potentials corresponds to the multi - soliton solutions of the two component CH equation.

The time evolution of $R_n$ due to (\ref{eq207}) is  

\b \label{eq40} R_n(t,\sigma)= R_n(0,\sigma)\exp\left(\frac{-\kappa_n}{\lambda_n(\sigma)}t
\right).  \e

As we already observed, both $\lambda_n(\sigma)$ and $R_n(t,\sigma)$ are
real. Let us define the $N\times N$ matrix 
\b M_{pq}(y,t,\sigma)=
\delta_{pq}-\sum_{n=1}
^{N}\frac{R_n(t,\sigma)R_q(t,-\sigma)e^{-2y(\kappa_n+\kappa_q)}}
{(\kappa_p+\kappa_n)
(\kappa_n+\kappa_q)},\label{Mpq}\e

which is real, and the vectors (with real components) \b A_n(y,t,\sigma)&=&
\sum_{p=1} ^{N}[M^{-1}(y,t,\sigma)]_{np},\label{An}\\
B_n(y,t,\sigma)&=&\sum_{p=1}
^{N}\sum_{q=1}^{N} [M^{-1}(y,t,\sigma)]_{np}\frac{R_q(t,\sigma)e^{-2y\kappa_q}}
{\kappa_p+\kappa_q}.\label{Bn}\e

The solution of the system (\ref{eq34}), (\ref{eq34a}) is \b \label{eq43a}
\underline{\psi}(x,-i\kappa_n,\sigma)&=& X_0(x)[A_n(y,t,\sigma)-e^{-2i\sigma
z}B_n(y,t,\sigma)],
\e

and from (\ref{eq33}) also

\b \label{psisoln}
\underline{\psi}(x,k,\sigma)&=&X_0(x)\Big[1-\sum_{n=1}
^{N}
\frac{iR_n(t,\sigma)e^{-2\kappa_n y}B_n(y,t,-\sigma)}{k-i\kappa_n}\nonumber
\\
&&\hspace{2mm}+\
e^{-2i\sigma z}\sum_{n=1} ^{N}
\frac{iR_n(t,\sigma)e^{-2\kappa_n y}A_n(y,t,-\sigma)}{k-i\kappa_n}\Big].\e

From (\ref{eq35}) we have

\b
e^{(-x+y)/2}&=&\cos(\sigma_1 z)X_0(x)\Big(1
- \sum_{n=1}^{N} \frac{R_n(\sigma_1)e^{-2\kappa_n y }
(A_n(y,t,-\sigma_1)-B_n(y,t,-\sigma_1))}{\kappa_n+1/2}
\nonumber \Big). \\  \label{eq:xy1}\e

A similar relation for $\underline{\varphi}(x,k,\sigma)$ can be written if $k\in \mathbb{C}_{+}$ and for  $k=i/2$ it gives

\b
e^{(x-y)/2}&=&\cos(\sigma_1 z)X_0(x)\Big(1
- \sum_{n=1}^{N} \frac{R_n(-\sigma_1)e^{-2\kappa_n y }
(A_n(y,t,\sigma_1)-B_n(y,t,\sigma_1))}{\kappa_n-1/2}
\nonumber \Big), \\  \label{eq:xy2}\e

and finally from (\ref{eq:xy1}) and (\ref{eq:xy2}):

\b x&\equiv & X(y,t)=  y+ \ln \frac{f_-(y,t)}{f_+(y,t)} \label{X} \,,\\
f_{\pm}(y,t)&=& 1
- \sum_{n=1}^{N} \frac{R_n(\pm\sigma_1)e^{-2\kappa_n y }
(A_n(y,t,\mp\sigma_1)-B_n(y,t,\mp\sigma_1))}{\kappa_n\pm1/2} \label{fpm}
\,.\e

The time evolution of $R_n$ is known (\ref{eq40}) and thus $X(y,t)$ is given
in terms of the scattering data. This
produces a parametric representation of the solution in terms of the dependent variables, from equations  (\ref{eq36}) -- (\ref{eq36a}).

For the one-soliton solution, equations (\ref{X}) and (\ref{fpm}) yield the parametric (Lagrangian) representation of the ``particle paths'',
\b \label{1sol} X(y,t)=y+\ln\frac{1-\frac{\frac{1}{2}+\kappa_1}
{\frac{1}{2}-\kappa_1}E(\sigma_1)E(-\sigma_1)+\frac{2\kappa_1}
{\frac{1}{2}-\kappa_1}E(-\sigma_1)}
{1-\frac{\frac{1}{2}-\kappa_1}{\frac{1}{2}+\kappa_1}E(\sigma_1)E(-\sigma_1)-\frac{2\kappa_1}
{\frac{1}{2}+\kappa_1}E(\sigma_1)},\e
where $E(\sigma,t)=\frac{1}{2\kappa_1}R_1(t,\sigma)e^{-2\kappa_1 y}$.

We notice that in (\ref{eq35}) one can take the imaginary part of the
eigenfunction which is also equal to $e^{-x/2}$, up to multiplicative constant
(since both the real and imaginary parts decay to zero as $x\to \infty$ and when $\lambda=0$ they both equal $e^{-x/2}$). The imaginary part yields another solution, which differs from the presented one by change of the signes of
the scattering data  $ R_i(t,\sigma)$.

Thus, the Inverse Scattering Transform method yields the parametric representation (\ref{X}) -- (\ref{fpm}) of the solution of the CH2 system in terms of the scattering data for its isospectral problem. From the viewpoint of fluid dynamics, this is a Lagrangian (particle) representation of the solution, which may be written in terms of the Eulerian (spatial) representation by using equations  (\ref{eq36}) -- (\ref{eq36a}).

%%%%%%%%%%%%%%%%%%%%%%%%%%%%%%%%%%%%%%%%%%%%%%
%      BEGIN REM
%%%%%%%%%%%%%%%%%%%%%%%%%%%%%%%%%%%%%%%%%%%%%%
\rem{
\subsection{Other examples from the CH2 hierarchy}

\subsubsection{CH2 system with two differential constraints}

With the same spectral problem, (\ref{L1}) and different (\ref{L2}) we can
obtain various members of the CH2 hierarchy. For example, if we substitute
(\ref{L2}) with  

\b  \Psi_{t}=\left(\frac{1}{2\lambda^2}-\frac{u_1}{\lambda}-u_0\right)\Psi_{x}+
\frac{1}{2}\left(\frac{u_{1,x}}{\lambda}+u_{0,x}\right)\Psi,\nonumber
\e
we obtain a system with differential constraints. They
 can be easily integrated,  giving

\b \label{L7} q_{1}&=&u_1-u_{1,xx}+\omega_1,\\
q_2&=&u_0-u_{0,xx}+3u_1^2-u_{1,x}^2-2u_1u_{1,xx}+4\omega_1
u_1+\omega_2,\label{L8}\e

where $\omega_{1,2}$ are arbitrary constants. The system of
equations for $u_0$, $u_1$ arising from the compatibility is  

\b  q_{2,t}\!\!&+&\!\!2u_{0,x}q_2+u_0q_{2,x}=0, \label{L9} \\
q_{1,t}\!\!&+&\!\!2u_{0,x}q_{1}+u_0q_{1,x}+2u_{1,x}q_{2}+u_1q_{2,x}=0.
\label{L10} \e

\begin{framed}
This system may be rewritten as 
\b
(\partial_t +\mathcal{L}_{u_0}) (q_2\,dx^2) = 0
,\qquad
(\partial_t +\mathcal{L}_{u_0}) (q_1\,dx^2) + \mathcal{L}_{u_1}\, (q_2\,dx^2)  = 0
\label{22-eq}
.\e
The system in (\ref{L9}-\ref{L10})  may appear to have two characteristic speeds, $u_0$ and $u_1$. However, as we shall see, its only characteristic speed is $u_0$. 

To understand system (\ref{22-eq}) geometrically, we first rederive it from a Lagrangian $l(u_0,q_2):\, \mathfrak{X}(\mathbb{R})\times \mathfrak{X}^*(\mathbb{R})\to \mathbb{R}$ using the Euler-Poincar\'e theory. Then we obtain its Hamiltonian formulation in terms of a Lie-Poisson bracket by performing a Legendre transformation. Here $\mathfrak{X}(\mathbb{R})$ is the space of real-valued right-invariant smooth vector fields on the real line $\mathbb{R}$, identified with 
\[
\mathfrak{X}(\mathbb{R})\simeq  T{\rm Diff}(\mathbb{R})/{\rm Diff}(\mathbb{R})
\]
and $\mathfrak{X}^*(\mathbb{R})$ is its dual space of real-valued smooth 1-form densities. Thus, the Lagrangian is right-invariant under ${\rm Diff}(\mathbb{R})$, the diffeomorphisms of the real line. We regard the dual space $\mathfrak{X}^*$ as simply a vector space on which the Lie algebra $\mathfrak{X}$ of vector fields acts by Lie derivative, e.g., as
\[
\mathcal{L}_{u_0} (q_2\,dx^2) =  (2u_{0,x}q_{2}+u_0q_{2,x})\,dx^2
\,,
\]  
which is a repeated pattern in the system (\ref{22-eq}). The Lie algebra action 
\[
\partial_t (q_2\,dx^2) =-\mathcal{L}_{u_0}(q_2\,dx^2)
\]
arises from a right Lie group action for the time evolution, denoted with subscript $t$ as
\[
(q_2\,dx^2)_t = (q_2\,dx^2)_0\circ g_t^{-1}
\quad\hbox{where}\quad 
\dot{g}_t=(u_0)_t\circ{g}_t
\,.\]

Calculating with Hamilton's principle $\delta S =0$ with $S=\int l(u_0,q_2)\,dt$ for this right-invariant Lagrangian as in \cite{HoMaRa1998,HoScSt2009} yields,
\begin{eqnarray*}
\delta \int_a^b l(u_0,q_2) dt 
&=&  \int_a^b 
\left\langle \frac{\delta l}{\delta {u}_0}, \delta{u}_0\right\rangle 
+
\left\langle \frac{\delta l}{\delta {q_2}}, \delta{q_2}\right\rangle
 dt
\\&=&  \int_a^b  
\left\langle  \frac{\delta l}{\delta {u}_0}
\,,\, 
\frac{d\eta}{dt} 
-
{\rm ad}_{u_0} {\eta}\right\rangle 
+
\left\langle \frac{\delta l}{\delta {q_2}}, -\mathcal{L}_\eta{q_2}\right\rangle
 dt
\\
&=&
 -\int_a^b\, 
 \left\langle  \frac{d}{dt} \frac{\delta l}{\delta {u_0}} 
+ 
{\rm ad}^*_{u_0} \frac{\delta l}{\delta {u_0}}
-
{\rm ad}^*_{\delta l/\delta {q_2}} q_2
\, ,\, {\eta} \right\rangle
\,dt \,,
\\
&=&
 -\int_a^b\, 
 \left\langle  \frac{d}{dt} \frac{\delta l}{\delta {u_0}} 
+ 
\mathcal{L}_{u_0} \frac{\delta l}{\delta {u_0}}
-
\mathcal{L}_{\delta l/\delta {q_2}} q_2
\, ,\, {\eta} \right\rangle
\,dt \,,
\end{eqnarray*}
so that Hamilton's principle implies the second equation in (\ref{22-eq}), upon identifying 
\begin{equation}
 \frac{\delta l}{\delta {u_0}}  = q_1
 \quad\hbox{and}\quad
   -\,\frac{\delta l}{\delta {q_2}}  = u_1
  \,.
  \label{q1u1-def}
\end{equation}

In this calculation, we have invoked $\eta=0$ and $\xi=0$ at the endpoints $a$ and $b$ in time when integrating by parts and have taken the continuum velocity vectors $u_0$ and $q_2$ to vanish asymptotically in space. We have also used the formulas
\begin{equation}
\delta{u_0}=
\frac{d\eta}{dt} 
-
{\rm ad}_{u_0} {\eta}
\,,\qquad
\delta{q_2}
=
-\mathcal{L}_\eta{q_2}
\,,
\quad\hbox{with}\quad
\eta \in \mathfrak{X}(\mathbb{R})
\,,
\label{right-invar-vars}
\end{equation}
for the variation of the right-invariant vector field $u$. We have taken 
$\langle\cdot\,,\,\cdot\rangle:\,\mathfrak{X}(\mathbb{R})\times \mathfrak{X}^*(\mathbb{R})\to \mathbb{R}$ as the $L^2$ pairing between elements of the Lie algebra $\mathfrak{X}(\mathbb{R})$ and its dual $\mathfrak{X}^*(\mathbb{R})$, the space of smooth 1-form densities. This pairing allows us to define the  ${\rm ad}^*$-operation as
\[
\left\langle 
 \frac{\delta l}{\delta {u_0}}\, ,\, {\rm ad}_{u_0}{\eta} \right\rangle
 =:
\left\langle 
{\rm ad}^*_{u_0} \frac{\delta l}{\delta {u_0}}\, ,\, {\eta} \right\rangle
 ,
 \]
where
\[
{\rm ad}_{u_0}{\eta} := - [u_0,\eta] := - u_0\partial_x\eta+ \eta \partial_x u_0
\]
is the adjoint Lie-algebra action $\mathfrak{X}(\mathbb{R})\times \mathfrak{X}(\mathbb{R})\to \mathfrak{X}(\mathbb{R})$ given by minus the Jacobi-Lie, or commutator bracket of smooth vector fields. 
In our case, taking $\langle\cdot\,,\,\cdot\rangle$ as the $L^2$ pairing between vector fields $\mathfrak{X}(\mathbb{R})$ and 1-form densities $\mathfrak{X}^*(\mathbb{R})$ yields
\[
\left\langle
\frac{\delta l}{\delta {u_0}}, \delta{u_0}\right\rangle
=
\int \delta {u_0}\,\frac{\delta l}{\delta {u_0}}\,dx
,
\]
in which integral is taken over the infinite real line. 
We may evaluate the ${\rm ad}^*$-expressions from their definitions, so that
\[
\left\langle {\rm ad}^*_{u_0} q_1\, ,\, {\eta} \right\rangle
=
\left\langle 
q_1\, ,\, {\rm ad}_{u_0}{\eta} \right\rangle
=
\left\langle 
q_1\, ,\, - v \eta_x+ \eta v_x \right\rangle
=
\left\langle \partial_x(q_1u_0) + q_1\partial_xu_0 ,\, \eta \right\rangle
=
\left\langle \mathcal{L}_{u_0} q_1\, ,\, {\eta} \right\rangle
.
\]
The $L^2$ pairing also allows us to formally identify $\mathfrak{X}^{**}(\mathbb{R}) \simeq  \mathfrak{X}(\mathbb{R})$. This identification and the previous line of manipulations produces the final term in the Hamilton's principle calculation above,
\[
\left\langle u_1\diamond {q_2}, \eta\right\rangle
:=
\left\langle u_1, -\mathcal{L}_\eta{q_2}\right\rangle
=
\left\langle u_1, -{\rm ad}^*_\eta{q_2}\right\rangle
\rem{
=
\left\langle -{\rm ad}_\eta u_1, {q_2}\right\rangle
=
\left\langle {\rm ad}_{u_1}\eta , {q_2}\right\rangle
}
=
\left\langle {\rm ad}^*_{u_1}{q_2},\,\eta \right\rangle
=
\left\langle \mathcal{L}_{u_1}{q_2},\,\eta \right\rangle
.\]
\paragraph{Legendre transformation.}
The Hamiltonian formulation of the system in (\ref{L9}-\ref{L10}) or  (\ref{22-eq}) is based on the  following Legendre transformation,
\[
h(q_1,q_2) = \left\langle q_1 , u_0 \right\rangle
-
l(u_0,q_2)
\]
with variations
\[
\delta h(q_1,q_2) = \left\langle \delta  q_1 , u_0 \right\rangle
+
\left\langle  q_1 -  \frac{\delta l}{\delta {u_0}} \,,\, \delta u_0 \right\rangle
+
\left\langle  u_1 \,,\, \delta q_2 \right\rangle
,\]
where in the last term, we have substituted the second relation in (\ref{q1u1-def}). Consequently, we may write the system in (\ref{L9}-\ref{L10}) or  (\ref{22-eq}) in Lie-Poisson Hamiltonian form, as
\begin{equation}
\begin{bmatrix}
\partial_t q_1 \\
\partial_t q_2
\end{bmatrix}
= - 
\begin{bmatrix}
\partial_x q_1 + q_1 \partial_x & \partial_x q_2 + q_2 \partial_x \\
\partial_x q_2 + q_2 \partial_x &  0
\end{bmatrix}
\begin{bmatrix}
\delta h /\delta q_1 = u_0 \\
\delta h /\delta q_2 = u_1
\end{bmatrix}
  \,.
  \label{SDP Ham form}
\end{equation}
The Hamiltonian operator yields the Lie-Poisson bracket dual to the semidirect product Lie algebra 
$\mathfrak{X}(\mathbb{R})\circledS\mathfrak{X}^*(\mathbb{R})$.  

\paragraph{Remark.}
If the problem specified here for a right-invariant Lagrangian on $\mathfrak{X}\times \mathfrak{X}^*$ had been expressed instead on $\mathfrak{so}(3)\times \mathfrak{so}(3)^*$ for a left-invariant Lagrangian, the result would have been the dynamics of a sort of rotating top in a potential force field. The corresponding dynamics would have been expressible as 
\begin{equation}
\begin{bmatrix}
\dot{q}_1 \\
\dot{q}_2
\end{bmatrix}
= 
\begin{bmatrix}
q_1\times & q_2\times \\
q_2\times &  0
\end{bmatrix}
\begin{bmatrix}
\delta h /\delta q_1 = u_0 \\
\delta h /\delta q_2 = u_1
\end{bmatrix}
  \,,
 \label{top-prob}
\end{equation}
for angular momenta $({q}_1,{q}_2)\in\mathbb{R}^3\times\mathbb{R}^3$
and corresponding angular velocities $(u_0,u_1)\in\mathbb{R}^3\times\mathbb{R}^3$.

\end{framed}

\subsubsection{The CH2 Dym equation }\label{CH2Dym}
There are members of CH2 hierarchy, for which the Lax pair contains only
\emph{positive} powers of $\lambda$. The simplest and most interesting example of this kind has a Lax pair
\b
\Psi_{xx}&=&\left(\lambda^2 \epsilon \rho^2+\lambda q + 1/4\right)\Psi, \\
\Psi_t&=& 2\frac{\epsilon \lambda}{\rho}\Psi_{x}-\left(\epsilon \lambda/\rho\right)_x\Psi
\,,
\label{CH2-1}
\e
with $\epsilon=\pm 1$. 
This coupled nonlinear system is
at the position in the CH2 hierarchy that corresponds to the modified Dym equation, first introduced as a tri-Hamiltonian system in \cite{SA}: 
%\begin{framed}
\begin{eqnarray}
\rho_t +\left(\frac{ q}{\rho^2} \right)_x &=& 0
\,,\label{rho-dot}\\
\ \epsilon q_t + \left(\big(1-\partial_x^2 \,\big)\frac{1}{\rho}\right)_x &=& 0
\,.\label{q-dot}
\end{eqnarray}
%\end{framed}
These equations may be written as compatibility conditions for the two differential 1-forms
\begin{eqnarray}
dp &=& \rho dx - \frac{q}{\rho^2}  \,dt
= p_x \,dx + p_t \,dt
\,,\label{dp-eqn}\\
dm &=& \epsilon q dx - (1-\partial_x^2)\frac{1}{\rho} \,dt
\,.
\end{eqnarray}
Thus, from (\ref{dp-eqn}) one identifies 
\begin{equation}
\rho = p_x
\,,\quad 
q = -\,p_x^2 p_t
\,.
\end{equation}
Consequently, $dm$ can be written in terms of $p$ as,
\begin{equation}
dm =  -\,\epsilon (p_t p_x^2)\, dx - \left((1-\partial_x^2)\frac{1}{p_x}\right)dt
\,.
\end{equation}
By requiring compatibility, one finds
\begin{equation}
d^2m = 0 = \left[ -\epsilon \partial_t(p_t p_x^2) 
+ 
\partial_x \left((1-\partial_x^2)\frac{1}{p_x}\right) \right] dt\wedge dx
\,,
\end{equation}
and this vanishing, i.e., 
\begin{framed}
\begin{equation}
-\epsilon \partial_t(p_t p_x^2) 
+ 
\partial_x \left((1-\partial_x^2)\frac{1}{p_x}\right) = 0
\,.
\label{p}\end{equation}
\end{framed}
Linearization of this equation around $p_t=0$ and $p_x=1$ yields a wave dispersion relation with real frequency for $p$, when $\epsilon =-1$.

\paragraph{Travelling wave solution.}
Equation (\ref{p}) has a travelling wave solution in terms of the variable $\xi=x-ct$, with  $c>0$ given  by
\b \frac{1}{p'}=c^{1/4}\cosh^{1/2}(2(\xi-\xi_0)), \qquad \xi_0=\text{const.},\e

or  \b p(\xi)=\frac{c^{-1/4}}{\sqrt{2}}F\left(\arcsin\sqrt{\frac{\cosh 2(\xi-\xi_0)-1}{\cosh 2(\xi-\xi_0)}},\frac{1}{\sqrt{2}}\right)+\text{const.}, \e where $F(x,k)$
is the Elliptic Integral of the first kind \cite{GR00}. The travelling wave solution $p(\xi)$ has the form
of a {\bfi dark soliton} with a minimum at $\xi_0$ and constant asymptotic behavior as $|\xi|\to \infty$.

\begin{figure}[t]
\begin{center}
\includegraphics*[width=0.475\textwidth]{rhodynamics-minus-L=80}
\includegraphics*[width=0.475\textwidth]{qdynamics-minus-L=80}
\end{center}\vspace{-5mm}
\caption{%\footnotesize
Results are shown for the evolution of the system (7.2-7.3) with $\epsilon=-1$ for $\rho$ (left) and $q$ (right), arising in a periodic domain of length $L=80$ from initial conditions that represent a dam-break $q(x,0)=\tanh((x-L_1)/\alpha)-\tanh((x-L_2)/\alpha),\quad \rho(x,0)=1$ with $\alpha=1$, $L_1=L/3$, $L_2=2 L/3$. Soliton solutions are seen to emerge and propagate in both directions. The collision process produces a slight refraction of the soliton trajectories. 
The same system with $\epsilon=+1$ leads to the spontaneous (and sudden) emergence of singularities.  
Figures are courtesy of V. Putkaradze.
\label{DymCH2-fig} }
\end{figure}

\paragraph{Hamiltonian form.}
The coupled system (\ref{rho-dot}-\ref{q-dot}) can also be written in Hamiltonian form, as
\begin{equation}
\partial_t 
\begin{bmatrix}
\rho \\
\epsilon q
\end{bmatrix}
=
\begin{bmatrix}
\partial  & 0 \\
0 & \partial-\partial^3
\end{bmatrix}
\begin{bmatrix}
-q/\rho^2 = \delta h / \delta \rho 
\\
1/\rho = \delta h / \delta q 
\end{bmatrix}
,\quad\hbox{with}\quad
h := \int (q/\rho)\,dx\,.
\end{equation}
The energy conservation law may be expressed for $\epsilon=-1$ in conservative form as 
\begin{equation}
\partial_t \left( \frac{q}{\rho} \right)
+ 
\partial_x \left(\frac{1}{2}\left(\frac{q}{\rho^2}\right)^2 
+
\frac{1}{2\rho^2}
-
\frac{1}{\rho}\partial_x^2\frac{1}{\rho}
+
\frac{1}{2}\left(\partial_x\frac{1}{\rho} \right)^2
\right)
=
0
\,.
\end{equation}
This is also a wave equation for $p$
\begin{equation}
\partial_t \left( p_tp_x \right)
+ 
\partial_x \left(\tfrac{1}{2}p_t^2 
+
\frac{1}{2p_x^2}
-
\frac{1}{p_x}\partial_x^2\frac{1}{p_x}
+
\frac{1}{2}\left(\partial_x\frac{1}{p_x} \right)^2
\right)
=
0
\,,
\end{equation}
and it takes an interesting form, particularly because it is quadratic. Some of its solution behavior is shown in Figure \ref{DymCH2-fig}

\subsubsection{Equations in the CH2 hierarchy with two time variables}\label{2timeCHeqns}
There is also an integrable CH2 system with two `time' variables ($t$ and $y$). In particular, consider the system%
\footnote{A two-time version of the CH equation has been considered previously in \cite{Iv2009}.}
\begin{framed}
The two-time CH2 system with $m=U_x-U_{xxx}$ is
\b 
m_t+2 U_{yx} m + (U_y+\gamma) m_x 
+ \rho 
\rho_y &=&0
, \label{m-eqn} \\
\rho_t + \Big((U_y +\gamma )\rho \Big)_x&=&0. 
\label{rho-eqn} 
\e

This system can be written equivalently in a fluids form as 
\b 
(m/\rho^2)_t + (U_y +\gamma )(m/\rho^2)_x&=& -\, \rho^{-1}\rho_y
, \label{m-eqn2} \\
\rho_t + \Big((U_y +\gamma )\,\rho \Big)_x&=&0, 
\label{rho-eqn2} 
\e
which shows that it has only one characteristic velocity, $dx/dt=(U_y +\gamma)$. 
\end{framed}

The two-time CH2 system can also be written as the compatibility condition for the following linear system ({\bfi Lax pair}) with a constant spectral parameter $\zeta$: 
\b 
\Psi_{xx}&=&\Big(-\zeta^2\rho^2+\zeta m+\frac{1}{4}\Big)\Psi
, \label{ev-prob}
\\
\Psi_{t}-\frac{1}{2\zeta}\Psi_y&=&-\,(U_y+\gamma)\Psi_x+\frac{1}{2}U_{yx}\Psi
.
\label{evol-y}
\e The first equation in this system is the spectral problem (\ref{L1}) of the CH2 hierarchy. The second equation introduces the other `time' derivative, with respect to $y$.  The system (\ref{m-eqn}),(\ref{rho-eqn}) appears on setting 
\[
\left(\partial_t - \frac{1}{2\zeta} \partial_y\right)\Psi_{xx}
=
\partial_x^2 \left( \Psi_{t}-\frac{1}{2\zeta}\Psi_y\right),
\]
then using (\ref{ev-prob}),(\ref{evol-y}) to eliminate higher derivatives and assuming $\zeta_t=0=\zeta_y$.

{\bf Remarks.}

$\bullet\quad$  The integrable system of two-time CH2 equations (\ref{m-eqn}),(\ref{rho-eqn}) reduces to CH2 for $x=y$ and $u=U_x$. 

$\bullet\quad$  Likewise, the special case $\gamma =0=\omega$ with initial condition $\rho=0$ admits $N$-peakon solutions, 
\b
m(x,t;y) = \sum_{a=1}^N p_a(t;y)\,\delta(x-q_a(t;y))
\,,
\label{2Dpeakons}
\e
with two `time' variables ($t$ and $y$).

\begin{figure}[ht]
\begin{center}
\includegraphics[scale=0.5,angle=0]{lake-james.jpg}
\end{center}
\caption{%\footnotesize
We are hoping for a figure from Lennon showing the dam-break solution behavior of the two-time CH2 equations (\ref{m-eqn}),(\ref{rho-eqn}), arising from initial conditions (\ref{dambreak-ic}) in a periodic domain. }
\label{Lennon_figure}
\end{figure}

\begin{framed}
Figure \ref{Lennon_figure} plots the evolution $(u,\rho)$ governed by equations (\ref{m-eqn}-\ref{rho-eqn}) in the periodic domain $\left[-L,L\right]$ with {\bfi dam-break initial conditions} given by
\begin{equation}
u\left(x,0\right)=0,\qquad{\rho}\left(x,0\right)
= 1 +  \tanh(x+a)-\tanh(x-a) 
\,,
\label{dambreak-ic}
\end{equation}
where $a\ll L$.

The dam-break involves a body of water of uniform depth, 
retained behind a barrier, in this case at $x=\pm a$.  If this barrier is
suddenly removed at $t=0$, then the water would flow downward and outward under gravity.  The problem
is to find the subsequent flow and determine the shape of the
free surface.  This question is addressed in the context of shallow-water
theory, e.g., by Acheson \cite{Ach1990}, and thus serves as a typical hydrodynamic
problem of relevance for the solutions.
\end{framed}
%%%%%%%%%%%%%%%%%%%%%%%%%%%%%%%%%%%%%%%%%%%%%
}%            END REM
%%%%%%%%%%%%%%%%%%%%%%%%%%%%%%%%%%%%%%%%%%%%%

\section{A modified version of CH2 arising from the 2D EPDiff equation}\label{EPDiff-CH-eqn-sec}

In this section we discuss a modified version of CH2 that does not lie within its integrable hierarchy but does admit peakon solutions. 
Namely, we consider solutions of the 2-component EPDiff equation that depend only on the first spatial variable $x\equiv x_1$ and do not depend on $x_2$: 
\b 
q_t\!\!&+&\!\!uq_x+2q u_x +\rho(1-\partial_x^2)^{-1} \rho_x=0
,\label{EPD1}\\
\rho_t\!\!&+&\!\!(u\rho)_x=0
.\label{EPD2}
\e

This system of equations has been considered previously \cite{HLT09} from another viewpoint and it is known
that it has peakon solutions. However, a totally different interpretation of these  solutions exists. Instead of two different types of variables $u$ and 
$\bar{\rho}:=(1-\partial_x^2)^{-1}\rho$, one a velocity and one an average density, we may imagine  
having two velocity components $u_1=u$ and $u_2=\bar{\rho}$.
With this interpretation, the modified CH2 equations (\ref{EPD1}),(\ref{EPD2}) in 1D are equivalent to the original EPDiff($H^1$) equation in 2D coordinates $(x_1,x_2)\in\mathbb{R}^2$,  \cite{HoMa2004}
\begin{equation}
\mathbf{m}_t + \mathbf{u}\cdot\nabla \mathbf{m} + (\nabla\mathbf{u})^T\cdot \mathbf{m} =0
\,,
\label{2D-EPDiff}
\end{equation}
with 2D momentum $\mathbf{m}=\mathbf{u}-\Delta\mathbf{u}=(m_1,m_2)^T$ and velocity $\mathbf{u}=(u_1,u_2)^T$ independent of the second coordinate, $x_2$.  Substitution of $\mathbf{m}(x_1,t)$ and $\mathbf{u}(x_1,t)$ into the 2D EPDiff($H^1$) equation (\ref{2D-EPDiff}) yields the 1D two-component system (\ref{EPD1}),(\ref{EPD2}) with $x=x_1$.

\begin{figure}[ht]
\begin{center}
\includegraphics[scale=0.4,angle=0]{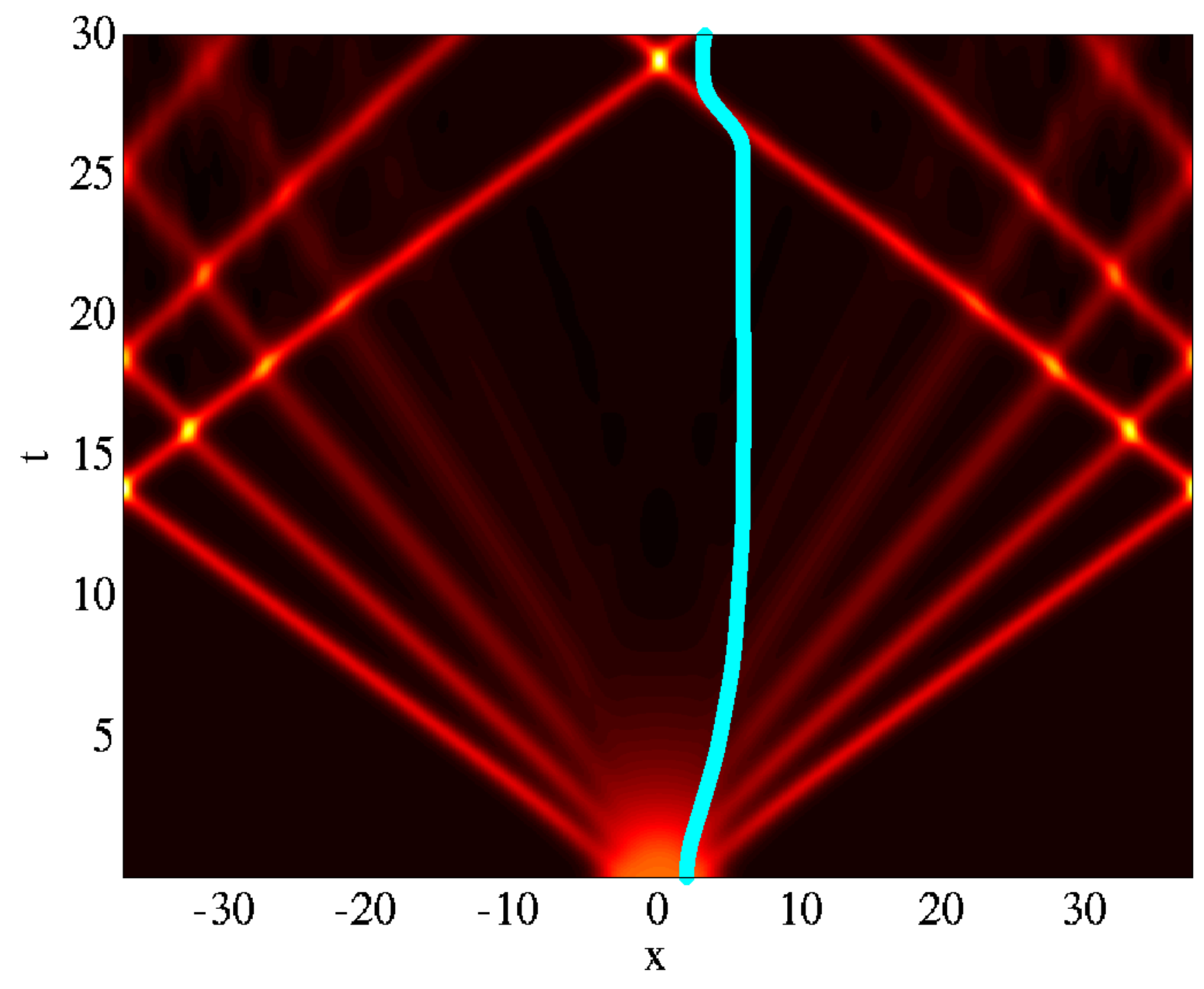}
\end{center}
\caption{%\footnotesize
The figure shows the particle path $x_1(t)$ for a single Lagrangian fluid parcel obtained from $dx_1/dt = u_1(x_1(t), t)$, upon identifying the modified CH2 solutions $(u,\bar{\rho})$ of the 1D dam-break experiments studied in \cite{HLT09} with solutions $(u_1,u_2)$ of the EPDiff$(H^1)$ equation (\ref{2D-EPDiff}) that are independent of the second coordinate, $x_2$, in a periodic domain. A fluid parcel initially offset to the right of center in $x_1$ accelerates gently to the right as the dam-break  produces pulses that propagate away on the right side. Later, when the leading pulse returns leftward, the parcel is pushed suddenly back with considerably greater leftward acceleration and ends up nearly where it started. Its trajectory shows that the wave pulses propagate relative to the Lagrangian parcels, so they are not frozen into the fluid motion. However, the influence of the returning pulse shows that the Lagrangian motion is still strongly coupled to the waves.  Figure courtesy of L. \'O N\'araigh.}
\label{peakon_figure} 
\end{figure}

The presence of peakon solutions of (\ref{EPD1}),(\ref{EPD2}) is not a surprise, since the singular
solutions are a characteristic feature for the EPDiff equation, which is
not known to be integrable beyond its one-dimensional version which coincides
with the CH equation. The peakon interactions can be studied numerically. An example is presented in Figure \ref{peakon_figure}, which shows a numerical simulation of the velocity $u_2(x_1,t)$ and the particle path $x_1(t)$ for a single Lagrangian fluid parcel obtained from $dx_1/dt = u_1(x_1(t), t)$, obtained as a solution of the EPDiff$(H^1)$ equation (\ref{2D-EPDiff}) in which the solutions $(u_1,u_2)$ are independent of the second coordinate.

%%%%%%%%%%%%%%%%%%%%%%%%%%%%%%%%%%%%%%%%%%%%%%%
%  BEGIN REM
\rem{
%%%%%%%%%%%%%%%%%%%%%%%%%%%%%%%%%%%%%%%%%%%%%%%

\subsection{A two-component cross-coupled generalization of CH2 with peakon solutions}

We consider a reduced Lagrangian $l({u},v):\,\mathfrak{X}(\mathbb{R})\times \mathfrak{X}(\mathbb{R})\to \mathbb{R}$ that depends on a pair of smooth vector fields $({u},v)\in \mathfrak{X}(\mathbb{R})=T{\rm Diff}(\mathbb{R})/{\rm Diff}(\mathbb{R})$ that are right-invariant under diffeomorphisms (smooth invertible maps) of the real line, ${\rm Diff}(\mathbb{R})$. Applying the standard calculation of Hamilton's principle for this right-invariant Lagrangian yields, see e.g.  \cite{HoMaRa1998,HoScSt2009},
\begin{eqnarray*}
\delta \int_a^b l({u},v) dt 
&=&  \int_a^b 
\left\langle \frac{\delta l}{\delta {u}}, \delta{u}\right\rangle 
+
\left\langle \frac{\delta l}{\delta {v}}, \delta{v}\right\rangle
 dt
\\&=&  \int_a^b  
\left\langle  \frac{\delta l}{\delta {u}}
\,,\, 
\frac{d\eta}{dt} 
-
{\rm ad}_{u} {\eta}\right\rangle 
+
\left\langle  \frac{\delta l}{\delta {v}}\,,\, \frac{d\xi}{dt} 
-
{\rm ad}_{v} {\xi}\right\rangle dt
\\
&=&
 -\int_a^b\, 
 \left\langle  \frac{d}{dt} \frac{\delta l}{\delta {u}} 
+ 
{\rm ad}^*_{u} \frac{\delta l}{\delta {u}}\, ,\, {\eta} \right\rangle
+
\left\langle  \frac{d}{dt} \frac{\delta l}{\delta {v}} 
+ 
{\rm ad}^*_{v} \frac{\delta l}{\delta {v}}\, ,\, {\xi} \right\rangle\,
\,dt \,,
\end{eqnarray*}
where, as in \cite{HoMaRa1998,HoScSt2009}, we have invoked $\eta=0$ and $\xi=0$ at the endpoints in time when integrating by parts and have taken the continuum velocity vectors $u$ and $v$ to vanish asymptotically in space. We have also used the formulas
\begin{equation}
\delta{u}=
\frac{d\eta}{dt} 
-
{\rm ad}_{u} {\eta}
\,,\qquad
\delta{v}
=
\frac{d\xi}{dt} 
-
{\rm ad}_{v} {\xi}
\,,
\quad\hbox{with}\quad
\eta,\ \xi \in \mathfrak{X}(\mathbb{R})
\,,
\label{right-invar-vars}
\end{equation}
for the variation of the right-invariant vector field $u$ and have taken 
$\langle\cdot\,,\,\cdot\rangle:\,\mathfrak{X}(\mathbb{R})\times \mathfrak{X}^*(\mathbb{R})\to \mathbb{R}$ as the pairing between elements of the Lie algebra $\mathfrak{X}(\mathbb{R})$ and its dual $\mathfrak{X}^*(\mathbb{R})$, the space of smooth 1-form densities. This pairing allows us to define the  ${\rm ad}^*$-operation as
\[
\left\langle 
{\rm ad}^*_{v} \frac{\delta l}{\delta {v}}\, ,\, {\xi} \right\rangle
=
\left\langle 
 \frac{\delta l}{\delta {v}}\, ,\, {\rm ad}_{v}{\xi} \right\rangle
 ,
 \]
where
\[
 {\rm ad}_{v}{\xi} := - [v,\xi] := - v\xi_x+ \xi v_x
\]
is the adjoint Lie-algebra action $\mathfrak{X}(\mathbb{R})\times \mathfrak{X}(\mathbb{R})\to \mathfrak{X}(\mathbb{R})$ given by minus the Jacobi-Lie, or commutator bracket of smooth vector fields. 
In our case, $\langle\cdot\,,\,\cdot\rangle$ is the $L^2$ pairing between vector fields $\mathfrak{X}(\mathbb{R})$ and 1-form densities $\mathfrak{X}^*(\mathbb{R})$, written as
\[
\left\langle
\frac{\delta l}{\delta {u}}, \delta{u}\right\rangle
=
\int \delta {u}\,\frac{\delta l}{\delta {u}}\,dx
,
\]
in which integral is taken over the infinite real line.

Substituting the definitions $m=\delta l/\delta {u}$  and $n=\delta l/\delta {v}$ 
produces a pair of Euler-Poincar\'e equations for coadjoint motion,
\begin{equation}
\partial_t m
+ 
{\rm ad}^*_{u} m
= 0
\,,\qquad
\partial_t n
+ 
{\rm ad}^*_{v} n
=
0
\,.
\label{cross-flow-eqns}
\end{equation}
We may evaluate the ${\rm ad}^*$-expressions from their definitions, so that
\[
\left\langle {\rm ad}^*_{v} n\, ,\, {\xi} \right\rangle
=
\left\langle 
n\, ,\, {\rm ad}_{v}{\xi} \right\rangle
=
\left\langle 
n\, ,\, - v\xi_x+ \xi v_x \right\rangle
=
\left\langle (nv)_x + nv_x ,\, \xi \right\rangle
.
\]  

Let us consider a few choices of the reduced Lagrangian $\ell({u},v)$ whose Euler-Poincar\'e equations support peakon solutions on the real line:
\begin{enumerate}
\item
When the reduced Lagrangian depends on only one vector field as $\ell({u})=\frac{1}{2}\|{u}\|_{H^1}^2$, then the dispersionless CH equation,
\[
\partial_t m
+ 
um_x + 2 mu_x
= 0
\,,
\quad\hbox{with}\quad
m = u - u_{xx}
,
\]
 emerges as the dynamics of geodesic motion on the diffeomorphisms with respect to the $H^1$ norm $\|{u}\|_{H^1}=\int u^2 + u_x^2\,dx$.
 \item
 When the reduced Lagrangian is taken as $\ell({u},v)=\int uv + u_xv_x\,dx$, we find a pair of coupled Euler-Poincar\'e equations as in  (\ref{cross-flow-eqns})
\begin{equation}
\partial_t m+ 2u_x m+u m_x= 0
\,,\qquad
\partial_t n + 2v_x n+ v n_x=0
\,,
\label{cross-eqns}
\end{equation}
 with $m=v-v_{xx}$ and $n=u-u_{xx}$. Because the momentum density ($m$) for one velocity vector field ($v$) is Lie-dragged by the vector field ($u$) for the other momentum density ($n$), we call this system the {\bfi cross-flow equations}. When $v=u$, this coupled system restricts to the dispersionless CH equation. (Dispersion may be introduced by introducing $m=v-v_{xx}+\omega_1$ and $n=u-u_{xx}+\omega_2$.)
  \item
When $(u,v)\in\mathbb{C}$ and $v=\bar{u}$, the reduced Lagrangian becomes the complex $H^1$ norm $\ell({u},v)=\int |u|^2 + |u_x|^2\,dx$ and we may interpret the coupled system of cross-flow equations (\ref{cross-flow-eqns}) as the complex version of the dispersionless CH equation. 
 \item
The coupled system of cross-flow equations (\ref{cross-flow-eqns}) two characteristic velocities. 
\end{enumerate}

\begin{figure}%[ht]
\begin{center}
\includegraphics*[width=0.475\textwidth]{xflow_u_revised}
\includegraphics*[width=0.475\textwidth]{xflow_v_revised}
\end{center}
\caption{\label{xflow-figs} This figure shows the dam-break solution behavior of the variables $v$ (left panel) and $u$ (right panel) of the cross-flow equations (\ref{cross-eqns}). The evolution may be compared with the corresponding results for the system of CH2 equations in (\ref{LL1})-(\ref{LQ1}), fig. \ref{CH2-figs}. Figures are courtesy of J. R. Percival.
} 
\end{figure}

\paragraph{``Dam-break'' equivalent problem for the cross-flow equations.}
The dam-break results for the cross-flow system in (\ref{cross-eqns}) in Figure \ref{xflow-figs} show evolution of the variables $v$ (left panel) and $u$ (right panel), arising from initial conditions similar to (\ref{dambreak-ic}) in a periodic domain, in which an initially localized disturbance with tanh-squared profile in $m$
interacts with an initially constant mean flow in the independent velocity field, $u$. That is,
\begin{align*}
m(x,0) & =\left[1+\tanh\left(x+1\right)-\tanh\left(x-1\right)\right]^{2},\\
n(x,0) & =u(x,0)=1.
\end{align*}
This dam-break initial condition first spawns a rapid leading pulse in velocity $u$ and then produces a series of slower pulses in $u$, some of which are of {\it larger amplitude} than the first pulse. The rapid first pulse in $u$ then passes again through the periodic domain, overtaking and colliding with the series of slower but larger pulses and undergoing a strong interaction indicated by a burst of amplitude in each collision. The $u$ pulse and the $v$ pulse track each other, but their variations have opposite phase. That is, a maximum in the $u$ pulse corresponds to a minimum in the $v$ pulse, and vice versa.  This is an interesting scenario that will be investigated further elsewhere. 

\paragraph{Peakon solutions of the cross-flow equations.}
The cross-flow equations (\ref{cross-flow-eqns}) for any of these choices of the reduced Lagrangian are deformations of CH that support two different types of peakons, with velocities
\b
v(x,t) =  \tfrac12\sum_{a=1}^M m_a(t)\,e^{-|x-q_a(t)|}
\,,\qquad
u(x,t) = \tfrac12\sum_{b=1}^N n_b(t)\,e^{-|x-r_b(t)|}
\,,
\label{xflow-peakons}
\e
and momenta,
\b
m(x,t) = \sum_{a=1}^M m_a(t)\,\delta(x-q_a(t))
\,,\qquad
n(x,t) = \sum_{b=1}^N n_b(t)\,\delta(x-r_b(t))
\,.
\label{xflow-peakons-veloc}
\e
The $2M+ 2N$ variables $(q_a,m_a)$, $a=1,\dots,M$, and $(r_b,n_b)$, $b=1,\dots,N$, are governed by the Hamilton's canonical equations for the Hamiltonian function,
\b
H = \tfrac12\sum_{a,b=1}^{M,N} m_a(t)n_b(t) e^{-|q_a(t)-r_b(t)|}
\,,\e
namely,
\b
\dot{q}_a(t)
&=& \frac{\partial H}{\partial m_a}
= \tfrac12 \sum_{b=1}^{N} n_b(t) e^{-|q_a(t)-r_b(t)|}
=u(q_a(t),t)
\,,\nonumber\\
\dot{r}_b(t)
&=& 
\frac{\partial H}{\partial n_b}
= \tfrac12 \sum_{a=1}^{M} m_a(t) e^{-|q_a(t)-r_b(t)|}
=v(r_b(t),t)
\,, \nonumber
\e
for the positions of the peakons, and
\b
\dot{m}_a(t)&=& 
-\,\frac{\partial H}{\partial q_a}
= \tfrac12 m_a \sum_{b=1}^{N}n_b\, {\rm sgn}\,(q_a-r_b)e^{-|q_a(t)-r_b(t)|}
= -\,m_a\,\frac{\partial u}{\partial x}\Big|_{x=q_a}
\,, \nonumber\\
\dot{n}_b(t)&=& -\,\frac{\partial H}{\partial r_b}
= \tfrac12 n_b \sum_{a=1}^{M} m_a\, {\rm sgn}\,(q_a-r_b)e^{-|q_a(t)-r_b(t)|}
= -\,n_b\,\frac{\partial v}{\partial x}\Big|_{x=r_b}
\,, \nonumber
\e
for their canonical momenta. For the case that $v=\bar{u}\in\mathbb{C}$ we have $N=M$ and $n_a=\bar{m}_a$. 
These peakon solutions will be studied and reported elsewhere. 

\paragraph{Analogy with coupled rigid body dynamics.}
If the problem specified here for a right-invariant Lagrangian on $\mathfrak{X}\times \mathfrak{X}$ had been expressed instead on $\mathfrak{so}(3)\times \mathfrak{so}(3)$ for a left-invariant Lagrangian, the result would have been interpretable as the dynamics of a coupled system of two rigid bodies. The corresponding dynamics would have been expressible in Lie-Poisson form and would be integrable as a Hamiltonian system, if an additional $SO(2)$ symmetry were present, as for the case of axisymmetric coupled rigid bodies. 

%%%%%%%%%%%%%%%%%%%%%%%%%%%%%%%%%%%%%%%%%%%%%%%
}
%  END REM
%%%%%%%%%%%%%%%%%%%%%%%%%%%%%%%%%%%%%%%%%%%%%%%

\section{Conclusions/Discussion} \label{conclusion-sec}

\paragraph{Main results of the paper.}
The stage was set in Section \ref{SpectralProblem-sec}  for our discussions of the inverse scattering transform method for the two-component CH2 system. The CH2 Jost solutions were obtained in Section \S\ref{sec:3} and their asymptotic behavior was usedin Section \S\ref{RHP-sec} to reformulate the scattering problem as a Riemann-Hilbert problem (RHP) . By solving the RHP, multi - soliton solutions of CH2  were obtained as reflectionless potentials in Section \S\ref{solitons}. The soliton solutions of CH2 arising from the RHP expressed themselves in a parametric form corresponding to the Lagrangian representation of fluid dynamics. A slightly modified version of CH2 was found in Section \S\ref{EPDiff-CH-eqn-sec} by considering translation invariant EPDiff solutions. The peakon solutions of EPDiff (diffeons) were shown graphically to interact with the Lagrangian fluid parcels by briefly sweeping them along the peakon trajectory. It remains an open problem, as to whether the modified version of CH2 in (\ref{EPD1}) -- (\ref{EPD2}) is integrable. 

\paragraph{Explicit outstanding problems.} 
The integrable systems properties CH2 discussed here open the door for further generalizations and applications, some of which have been presented in the paper and others that will be discussed elsewhere. In particular, CH2 is a member of a large family of integrable multi-component PDE based on the Schr\"odinger equation with an energy-dependent potential, as discussed in \cite{HoIv2010}. The numerical simulation of these PDE, and the formulation and analysis of their discrete integrable versions can be expected to attract considerable attention in future endeavors. One may expect the continuing interest in wave-breaking analysis for CH and CH2 to apply to the integrable PDE in the rest of the CH2 family, as well.

\section*{Acknowledgements}
DDH was partially supported by the Royal Society of London, Wolfson Scheme.
RII acknowledges funding from a Marie Curie Intra-European Fellowship. 
Both authors thank L. \'O N\'araigh and J. R. Percival for generously providing figures from their numerical solutions in ongoing investigations of the various equations treated here.

\end{document}